\documentclass[prd,twocolumn,showpacs,superscriptaddress,dblfloatfix,preprintnumber,floatfix,footinbib]{revtex4-2}

\usepackage{graphicx} 
\pdfoutput=1

\usepackage{amsmath,amssymb}
\usepackage{epsfig}
\usepackage{hyperref}
\usepackage[dvipsnames]{xcolor}
\usepackage{slashed}
\usepackage{placeins}

\usepackage{amsfonts}
\usepackage{graphicx, rotating}
\usepackage{epstopdf}
\usepackage{latexsym}
\usepackage{rotating}
\usepackage{braket}
\usepackage{multirow}

\usepackage[export]{adjustbox}
\usepackage[version=3]{mhchem}

\usepackage[font={small}]{caption}   

\definecolor{myred}{rgb}{0.7, 0, 0}
\definecolor{myblue}{rgb}{0, 0, 0.7}
\definecolor{mygreen}{rgb}{0.04, 0.7, 0.5}
\definecolor{mygray}{rgb}{0.1, 0.1, 0.1}

\hypersetup{colorlinks,citecolor=myred,linkcolor=myblue,urlcolor=myblue,linktocpage=true}

\def\be   {\begin{equation}}   \def\ee   {\end{equation}}
\def\ba   {\begin{array}}      \def\ea   {\end{array}}
\def\bea  {\begin{eqnarray}}   \def\eea  {\end{eqnarray}}
\def\bean {\begin{eqnarray*}}  \def\eean {\end{eqnarray*}}
\def\nn{\nonumber}
\def\bry{\begin{array}}
	\def\ery{\end{array}}

\def\GeV{\,{\rm GeV}}
\def\MeV{\,{\rm MeV}}
\def\keV{\,{\rm keV}}
\def\eV{\,{\rm eV}}

\def\meV{\,{\rm meV}}

\newcommand{\skipnew}[1]{}
\newcommand{\AB}[1]{{\color{cyan}[AB: #1]}}

\newcommand{\Moss}{M\"ossbauer\,\,}

\numberwithin{equation}{section}

\def\UMD{\small{Maryland Center for Fundamental Physics, University of Maryland, College Park, MD 20742, USA}}

\usepackage{tikz,xcolor,hyperref}
\definecolor{lime}{HTML}{A6CE39}
\DeclareRobustCommand{\orcidicon}{%
	\begin{tikzpicture}
		\draw[lime, fill=lime] (0,0) 
		circle [radius=0.16] 
		node[white] {{\fontfamily{qag}\selectfont \tiny ID}};	\draw[white, fill=white] (-0.0625,0.095) 
		circle [radius=0.007];	\end{tikzpicture}
	\hspace{-2mm}}
\foreach \x in {A,...,Z}{%
	\expandafter\xdef\csname orcid\x\endcsname{\noexpand\href{https://orcid.org/\csname orcidauthor\x\endcsname}{\noexpand\orcidicon}}
}

\date{\today}

\begin{document}
	\title{Probing (Ultra-) Light Dark Matter Using Synchrotron Based \Moss Spectroscopy}
	\author{Abhishek Banerjee\orcidA{}\,}
	\email{abanerj4@umd.edu}
	\affiliation{\UMD}
	
	\begin{abstract}
		
		We propose to search for (ultra)-light scalar dark matter (DM) using synchrotron radiation based \Moss spectroscopy technique. 
  Such DM induces temporal variation in various fundamental constants, which in turn causes time modulation of the nuclear transition energies. 
        When a \Moss source and absorber is separated by a large baseline, 
	the DM 
 induced 
 shift between their energy levels 
 can be 
    tested by the modulation of the photon absorption spectrum. 
        The narrow \Moss transitions 
        allow the setup to efficiently probe DM in the high frequency range. 
        We show that the reach of a \Moss experiment with the existing synchrotron beams is at par with the bounds from various equivalence principle violation searches. 
        An improvement of the synchrotron setup would enable us to
        probe the  hitherto uncharted territory of the DM parameter space upto MHz frequency. The proposed method would extend the DM search beyond the EP limit by several orders of magnitude,  and would provide the best bound on DM interaction strength with various standard model fields in the high ($>$kHz) frequency region.  
        

        
		
	\end{abstract}
	
	\maketitle

	\section{Introduction}

	Unveiling the nature of dark matter (DM), which accounts for 25\% of the energy budget of the universe, is one of the main quests of modern physics. 
	Ultra-light (sub-eV)  bosonic fields are natural dark matter candidates \cite{Kolb:1990vq}. They emerge in a number of theories aimed at solving outstanding problems of particle physics such as the strong CP  \cite{Peccei:1977ur,Peccei:1977hh,Wilczek:1977pj,Kim:1979if,Shifman:1979if,Zhitnitsky:1980tq,Weinberg:1977ma,Dine:1981rt,Dine:1982ah}  and hierarchy problems \cite{Adelberger:2003zx, Graham:2015cka, Flacke:2016szy,Banerjee:2020kww,Hook:2018jle}. If such particles exist, due to the misalignment mechanism, they are generically expected to have a cosmic abundance which can easily be all of the dark matter ~\cite{Abbott:1982af,Preskill:1982cy,Banerjee:2018xmn}. These strong theory motivations have sparked a significant experimental program aimed at discovering a broad range of such interactions \cite{Kim:2022ype,Blum:2014vsa,Banerjee:2022sqg,Brzeminski:2020uhm}. 
	
	The key idea behind these experiments is as follows. 
 The energy density in ultra-light dark matter (ULDM) is contained in coherent oscillations of the classical field $\phi\left(t, \vec{x}\right)$ associated with the dark matter particle ($\phi$). The oscillations occur at a frequency equal to the mass $m_{\phi}$ of the dark matter, with a width $\sim 10^{-6} m_{\phi}$ determined by the kinetic energy of the dark matter in the galaxy. When these particles interact with the standard model (SM), these coherent oscillations lead to time dependent signatures over a narrow frequency band.  These narrow band signatures are the prime targets of these experiments. 
	
	One of the generic kinds of interactions that these particles can have with the standard model is via CP-even moduli-like interactions where the oscillating dark matter field results in oscillations in various fundamental constants such as the fine structure constant, quark and lepton masses and the QCD scale. These kinds of dark matter interactions can be probed by looking for frequency comparisons between two sources of stable frequencies, typically realized via a suitable atomic clock system \cite{RevModPhys.90.025008,Antypas:2019yvv, Arvanitaki:2014faa, Antypas:2022asj}. In such systems, the time needed to drive the requisite clock (or interferometric) transitions limits the high frequency ($>$ kHz) reach, despite their exquisite sensitivity \cite{Filzinger:2023zrs,Banerjee:2023bjc,PhysRevLett.125.201302, PhysRevLett.129.241301}. Moreover, atomic clock systems are sensitive to electronic levels and thus have reduced sensitivity to physics that affects nuclear parameters. Given these challenges, it is interesting to develop techniques that can broaden the search for such physics. 
	
	In this work, we propose the use of \Moss spectroscopy to probe ultralight dark matter in the  kHz-MHz mass range. 
 \Moss spectroscopy is an age old technique, based on the recoil free resonant absorption of the emitted photon by a nucleus in a solid~\cite{mossbauer1958kernresonanzfluoreszenz,mossbauer1959kernresonanzabsorption,mossbauer2000discovery}. 
Usually the set up consists of a lattice of some specific radio nuclei which emits photons (emitter), and another lattice with the same nuclei in the ground state to absorb (absorber) the emitted photons. 
The photon absorption spectrum is tested in a spectroscope (see~\cite{gutlich2012mossbauer,nasu2012general,greenwood2012mossbauer} and refs. therein). 
To probe DM, we take a \Moss emitter and absorber and separate them by a baseline. The dark matter background alters the nuclear transition energy of the emitter and the absorber. When the \Moss emitter produces a photon, in the time taken by the photon to traverse the baseline and reach the absorber, the oscillating dark matter field changes the transition energy of the absorber relative to the transition energy of the emitter, leading to modulations in the absorption spectrum of the \Moss spectroscope. The short decay times of the \Moss nuclei permits this setup to probe frequencies higher than possible with conventional atomic setups and the system is directly sensitive to physics that changes nuclear properties. To maximize the sensitivity of this setup, it is desirable to have as large a baseline as possible between the emitter and the absorber. In conventional \Moss spectroscopy, where the emitter is a suitable radioactive source, the produced photons are not collimated. This limits the achievable baseline. However, with the advent of powerful synchrotron sources, it has now become possible to realize \Moss spectroscopy using such sources. In this case, the synchrotron source coherently drives a \Moss emitter, resulting in all the nuclei in the sample being driven to the corresponding excited states in phase. The subsequent decay of these nuclei also occurs in phase resulting in an enhanced collimated beam being produced in the forward direction, enabling $\sim 100 $ m - km scale baselines in such setups without loss of flux. For this reason, we will focus our attention on \Moss spectroscopy in synchrotron sources. 
	
	The rest of this paper is organized as follows. In Section~\ref{sec:exp_theory} and Section~\ref{sec:exp_theory_2}, we discuss the setup for the experiment and describe the main theoretical and experimental factors. 
	In Section ~\ref{sec:tech}, we discuss the parameters of synchrotron sources that are presently achievable. Using these, in Section ~\ref{sec:sensitivity}, we estimate the reach for various dark matter - standard model interactions. We describe the results in Section~\ref{sec:reults_discussions} and

	\section{Setup}\label{sec:exp_theory}
	
\begin{figure}[t!]
		\centering
		\includegraphics[width=\columnwidth]
		{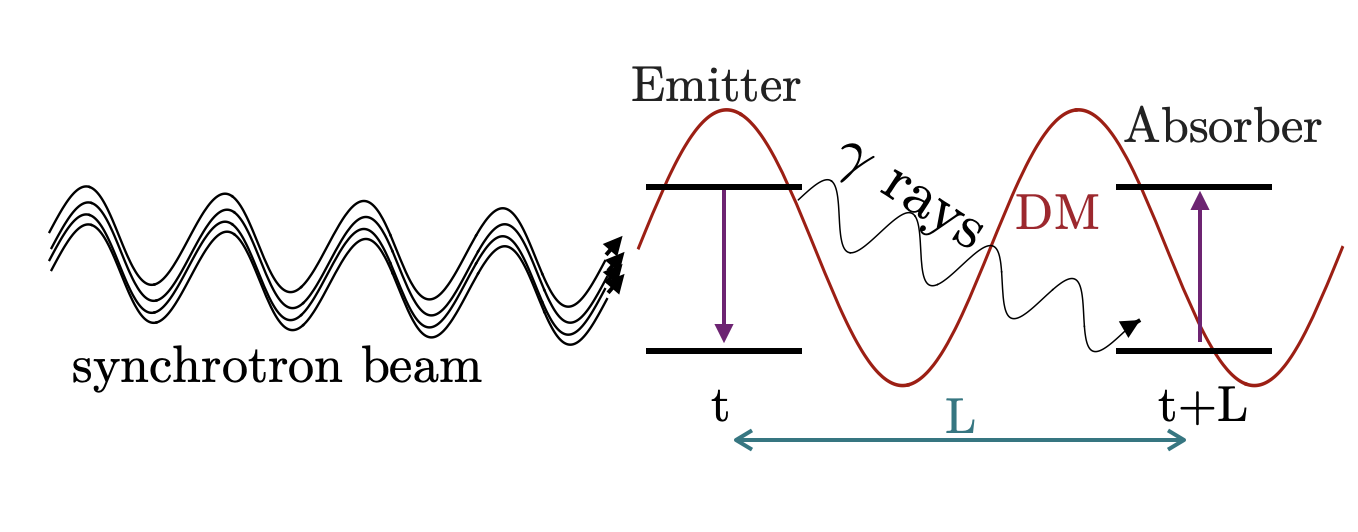}
		\caption{Proposed set up of the experiment.  
		}
		\label{fig:setup}
	\end{figure}

	Ultra-light bosonic dark matter is described in the galaxy as an oscillating classical field, with the oscillations centered around the mass $m_{\phi}$ of the particle with a width $10^{-6} m_{\phi}$. For concreteness, we will take the example of spin 0 fields, although the described phenomenology can also be extended to vectors~\footnote{Note that the sensitivity of the proposed setup to vector interactions is typically suppressed, unlike the case for scalar and tensor interactions.} and tensors.  If this field is the dark matter, its value in the galaxy is given by: 
	
	\bea
	\phi(t,\vec x)\simeq\frac{\sqrt{2 \rho_{\rm DM}}}{m_\phi}\cos\left[m_\phi (t+\vec\beta\cdot\vec x)\right]\,,
	\label{eq:phi_bg}
	\eea
	where, the oscillation amplitude is fixed by the requirement that $\phi$ accounts for 100\% of the dark matter (DM)  density $\rho_{\text{DM}} \sim 0.4\, \text{GeV/(cm)}^3$, and $|\beta|\sim 10^{-3}$ is the virial velocity. To see how such a dark matter field can give rise to a signal in a \Moss spectrometer, assume that the nuclear transition energy $E_{\gamma}$ is proportional to $\phi$, $E_{\gamma} = \kappa \phi$ (see section \ref{sec:sensitivity} for specific examples). When this is the case, suppose the emitter (see Figure~\ref{fig:setup})
	produces a photon at time $t$. The energy of the produced photon is $E_{\gamma} = \kappa \phi\left(t\right)$. This photon traverses a baseline of length $L$ when it encounters an absorber. The transition energy at the absorber is $E_{\gamma} = \kappa  \phi\left(t + L\right)$. This results in a difference between the emitter and absorber transition energies  equal to
	\bea
	\!\!\!\!\!\!\!\!\!\!\!
	\Delta E =  \kappa \frac{\sqrt{2 \rho_{\rm DM}}}{m_{\phi}/2} \sin\left( \frac{m_{\phi} L}{2}\right) \sin \left(\!m_{\phi}\!\left[t + \frac{L}{2}\right]\!\right)\!. 
	\label{Eqn:MainShift}
	\eea
	The \Moss spectrometer can be used to detect this shift with high accuracy. Note that for \eqref{Eqn:MainShift} to be valid,  the dark matter mass $m_{\phi} \lesssim \Gamma$ where $\Gamma$ is the decay width of the \Moss transition. This is because the decaying nucleus takes a time $\sim 1/\Gamma$ to produce the photon and modulations at frequencies higher than $\Gamma$ result in the production of sidebands around the transition energy as opposed to a homogeneous shift to the produced energy. The latter is the case of interest here and it occurs in the limit $m_{\phi} \lesssim  \Gamma$. This signal clearly oscillates at the frequency $m_{\phi}$, the smoking gun signature of ultra-light dark matter (ULDM). 
 
 For a qualitative understanding of the signal strength, we can recast Eq.~\eqref{Eqn:MainShift} as,
\bea
\!\!\!\!\!\!\!\!\!\!\!
\Delta E\propto  \! \sqrt{2 \rho_{\rm DM}} L\!\times\! \mathcal{W}(m_\phi L/2)\!\times
\!\sin\!\!\left[m_\phi\!\left(\!t+\frac{L}{2}\!\right)\!\right]\!,
\eea
where, $\mathcal{W}(x)={\rm sinc} (x)$. 
Notice, $\mathcal{W}(x)\simeq 1$ for very small $x$, and bounded as $|\mathcal{W}(x)|\lesssim x^{-1}$ for $x\gg 1$.
Thus, for a given (non zero) DM mass, the signal is maximized when the baseline $L \sim 1/m_{\phi}$. 
Note that, if the  total integration time $t_{\rm int}$ is longer than the DM coherence time $\tau_{\rm coh}=2\pi/(m_\phi\beta^2)$, the sensitivity of the experiment improves mildly as  $t_{\rm int}^{1/4}$~\cite{Budker:2013hfa,Derevianko:2016vpm}. 
Thus, for a given integration time, maximum sensitivity is reached at $m_\phi= 2\pi/(t_{\rm int}\beta^2)$.

\skipnew{
scales as $(t_{\rm int}\tau_{\rm coh})^{-1/4}$. 
due to the time-oscillating part, if the integration time of the experiment $t_{\rm int}$ is longer than the DM coherence time $\tau_{\rm coh}=2\pi/(m_\phi\beta^2)$, the sensitivity scales as $(t_{\rm int}\tau_{\rm coh})^{-1/4}$.  
}

\skipnew{
at a mild DM mass dependence is received for $m_\$
where, $t_{}$receives a mild ($m_\phi^{1/4}$) dependence on the DM mass. 

dependence of  $(t_{\rm int}\tau_{\rm coh})^{-1/4}$

For larger DM mass, The measurement time is longer than signal coherence time: T > τ

We have the usual scaling Bs ∝ T−1=2 as long as the
signal is phase coherent, 
Beyond the coherence time, the
sensitivity of the experiment scales as 

scales as T−1=4

which depends on $L$. 
		Note that, to obtain the DM induced energy shift, $(\Delta E)_{\rm DM}$, we need to specify its interaction with the SM, which is discussed in Section~\ref{sec:sensitivity}. 
		For a qualitative understanding of the signal strength, note that $\mathcal{W}(x)\simeq 1$ for very small $x$, and for $x\gg 1$, it is bounded as $|\mathcal{W}(x)|\lesssim x^{-1}$.  
		Thus, for $m_\phi$ up-to the order of the inverse of the distance between the emitter and the absorber, $L$, the signal remains independent of $m_\phi$, and after that it decreases as $1/m_\phi$.

note that the DM induced energy shift $\Delta E\propto \Delta\phi\simeq\phi(t)-\phi(t+L)$. 
For a baseline of $L$, the change of the DM field is

is proportional to the change of the DM field $\Delta \phi\simeq\phi(t)-\phi(t+L) \simeq \frac{\sqrt{2 \rho_{\rm DM}}}{m_\phi/2}\sin\left(\frac{m_\phi\,L}{2}\right)$

{\color{red}
		In the context of DM detection, due to its oscillating background,  
		DM induces changes to the nuclear transition energy at the \Moss emitter and the absorber at two different times. 
		For a spatial separation of $L$ between the emitter and the absorber, the net energy shift between them would be proportional to the change of the DM field amplitude change as,
		\bea
		\!\!\!\!\!\!\!\!\!\!
		\Delta \phi\simeq\phi(t)-\phi(t+L) \simeq \frac{\sqrt{2 \rho_{\rm DM}}}{m_\phi/2}\sin\left(\frac{m_\phi\,L}{2}\right),
		\eea
		where, we have omitted a factor of $\sin[m_\phi (t+L/2)]\sim \mathcal{O}(1)$ with $t$ being some initial time. 
		We recast the above equation as
		\bea
		\Delta \phi\simeq \sqrt{2 \rho_{\rm DM}}\, L\times \mathcal{W}(m_\phi L/2) \,,
		\label{Eq:del_phi}
		\eea
		where, $\mathcal{W}(x)=\sin(x)/x$ is the sinc function which depends on $L$. 
		Note that, to obtain the DM induced energy shift, $(\Delta E)_{\rm DM}$, we need to specify its interaction with the SM, which is discussed in Section~\ref{sec:sensitivity}. 
		For a qualitative understanding of the signal strength, note that $\mathcal{W}(x)\simeq 1$ for very small $x$, and for $x\gg 1$, it is bounded as $|\mathcal{W}(x)|\lesssim x^{-1}$.  
		Thus, for $m_\phi$ up-to the order of the inverse of the distance between the emitter and the absorber, $L$, the signal remains independent of $m_\phi$, and after that it decreases as $1/m_\phi$. 
		This allows us to efficiently probe DM with masses much larger than the ones typically probed by the frequency comparison tests.
 }
 }

	\section{Theoretical and Experimental Considerations}\label{sec:exp_theory_2}

	In this section, we outline the set up for the experiment and describe the main theoretical and experimental
	factors. 
	
	\Moss spectroscopy relies on the recoil free resonant absorption of the emitted gamma ray by a nucleus, known as the \Moss effect. 
	This is typically observed in the lattice of some specific radio nuclei~\cite{mossbauer1958kernresonanzfluoreszenz,mossbauer1959kernresonanzabsorption,mossbauer2000discovery}. 
	For a resonance of energy $E_0$, and a natural line width $\Gamma$, the absorption cross-section $\sigma_{\rm res}$, of a photon of energy $E_\gamma$ can be written as~\cite{laser},
	\bea
	\!\!\!\!\!\!\!\!\!\!\!\!\!\!\!
	\sigma_{\rm res}(E_0,E_\gamma,\Gamma)\propto \frac{2\pi}{E_\gamma^2}\frac{(\Gamma/2)^2}{(E_0-E_\gamma)^2+(\Gamma/2)^2}\,,
	\eea
	for a narrow resonance i.e.  $\Gamma/E_0\ll 1$~\footnote{We have omitted the factors involving the spins of the transition states, and the internal conversion factors~\cite{Wolfgang_Sturhahn_2004}.}. 
	As the \Moss transitions are very narrow, an energy shift of $\mathcal{O}(\Gamma)$ by the any perturbation would lead to an ``off resonance" transition, and a significant change in the  absorption cross section can be observed. 
 This unique property makes it an exciting probe not only for DM but also for new physics discoveries in general~\cite{Gratta:2020hyp}. 
	\skipnew{A typical \Moss transition has the energy of $\mathcal{O}(5-100)\keV$, and the width of $\mathcal{O}(0.1-1000)\,{\rm neV}$. 
		This $\mathcal{O}(10^{13})$ of magnitude difference in the energy scales involve in the process, makes it very susceptible to any environmental perturbations. 
		A detailed analysis of various background effects was done in~\cite{Gratta:2020hyp}, in the context of perturbing the \Moss set-up by introducing a attractor either very close to the emitter, or to the receiver. 
		It was shown that the most dominant background is due to the electromagnetic interactions.
		
		Nonetheless, the shielding effect due to the electron, and the suppression of the nuclear moments by the nucleus size, result in minimal influence on nuclear energy shifts from such interactions.
	}\\
	 
		


In our case, probing DM involves testing the photon absorption spectrum 
due to the energy level difference of the \Moss source and absorber. 
Thus, the experimental sensitivity depends on the photon flux at the absorber, which originates from the decay of the \Moss source. 
For a flux of $N_\gamma$ at the \Moss absorber, we obtain the sensitivity of the experiment as
\bea
	\Delta E = \frac{\Gamma}{\sqrt{N_\gamma}} 
	\,.
	\label{Eq:deltaE_exp}
	\eea
As the source and the absorber are kept close to each other, in conventional \Moss spectroscopy, the received flux is equal to the number of observed decays. 
\\

\skipnew{
on the number of observed  
of coherent decays
depends on the number of nuclei that participated in the initial emission. 
	For $N_\gamma$ number of coherent decays, the estimated total energy shift can be written as, 
	\bea
	\Delta E = \frac{\Gamma}{\sqrt{N_\gamma}} 
	\,.
	\label{Eq:deltaE_exp}
	\eea 
	We note that, for some transitions, such as \ce{^{181}Ta}, the observed width is broadened compare to the natural one~\cite{DORNOW1979491}. 
	In such case, one should use the observed width to estimate the sensitivity. {\color{red}
	As an estimate, consider a \Moss nuclei with   
    a decay time $\tau_{\rm dec}$ of $100\,{\rm ns}$, and $N_\gamma\sim 10^{13}$. 
	The energy shift can be obtained as,
	\bea
	\!\!\!\!\!\!\!\!\!\!\!\!\!\!
	\Delta E \simeq 2\times 10^{-15}\eV \left(\frac{100\,{\rm ns}}{\tau_{\rm dec}}\right)\left(\frac{10^{13}}{N_\gamma}\right)^{1/2}\!.
	\label{Eq:deltaE_estimate}
	\eea }
}

	One of the challenges of our proposed method is to maintain a high flux of photons even 
when the \Moss emitter and the absorber are far apart. 
	The effective reach of the experiment depends on 
 the number of received photons $N_\gamma$, and $L$, as $\sqrt{N_\gamma} \times L$. 
	Thus a decreasing flux $N_\gamma \propto 1/L^2$~\cite{Wolfgang_Sturhahn_2004}, mitigates the apparent gain in sensitivity obtained by a large baseline. 
	At the same time, if the emitted photons are not well collimated (which is the case for a traditional \Moss source), 
 the beam diverges quickly; results in an even smaller effective flux at the receiver.   
	\\
	
	To alleviate the challenges mentioned above, one possibility is to consider synchrotron radiation (SR) based \Moss spectroscopy~\cite{PhysRevLett.66.770,PhysRevLett.54.835,10.1007/978-94-010-0045-1_2,Seto2017-yy,Chumakov1999-ry,Yoda2019-qj,Smirnov1999-go}. 
 We will denote this set up as synchrotron based \Moss spectroscopy or SMS. 
	In SMS, a sample (placed $\sim \mathcal{O}(50\,{\rm m})$ away from the synchrotron source) is directly excited 
	through the absorption of an x-ray photon. 
	In contrast, in conventional \Moss spectroscopy, a radioactive parent undergoes decay to generate the desired isotope in the excited state. 

\skipnew{{\color{red}There are several advantages of using SMS; one of them has to do with the fact that in SMS, the source and receiver can be made out of the same material leading to the resonant absorption of the emitted photon is observed automatically. 
			Thus, probing the DM induced shift in the nuclear energy would require minimum amount of scanning of the line-width. }}

In SMS, a synchrotron source coherently excites all the nuclei in the \Moss emitter. 
 Thus the subsequent decay of these nuclei also happens in phase in the sense of superradiance~\cite{Chumakov2018-pg}.   
 This results in an enhanced collimated beam in the forward direction with a flux equal to the number of resonantly excited \Moss source nuclei. 
 \skipnew{
 drives the

resulting
in all the nuclei in the sample being driven to
the corresponding excited states in phase. The
subsequent decay of these nuclei also occurs in phase
resulting

 Thus, the photon emission happens coherently from the whole \Moss source, and the flux is equal to the number of source nuclei that has been excited by the SR.

 However, the most relevant benefit of the set-up stems from the fact that it is based on coherent elastic scattering of the x-ray~\cite{Smirnov1999-go}. 
		Thus, not only the initial and final states of the lattice states stay the same (coherent recoil of the whole lattice), but also the final and the initial state of the nuclei participating in the scattering process. 
		This means the emitted radiation is coherent, as if which atom scattered the photon cannot be decided. 
		This leads to an enhanced collimated emitted x-ray photons with an effective surface area of $4\pi$ in square radians. 
		This result is experimentally shown in~\cite{Chumakov2018-pg}. 
		Thus, the photon emission happens coherently from the whole \Moss source, and the flux is equal to the number of source nuclei that has been excited by the SR. }
	However, as the resolution of the synchrotron source, $(\Delta E)_{\rm SR}$, is much larger than the extremely narrow \Moss transitions, 
	only a fraction of the \Moss source nuclei would be resonantly excited. 
	Thus, the 
 emitted flux at the \Moss source  $(N_{\gamma})_{\rm MS}$, can be calculated as,
	\bea
	(N_{\gamma})_{\rm MS} = (N_{\gamma})_{\rm SR}\frac{\Gamma}{(\Delta E)_{\rm SR}}\,,
	\label{eq:fulx_moss_source}
	\eea
	where, $(N_{\gamma})_{\rm SR}$ is the incident synchrotron flux at the \Moss source. 
	\\
	
	To achieve a large statistics, we need to make sure that the flux at the \Moss absorber is not significantly reduced. 
	To estimate this, we calculate the Rayleigh range  $z_R$, of a beam. Qualitatively a beam   remains parallel over twice the Rayleigh range~\cite{laser}. 
	To estimate the flux at a distance $z$, $N_\gamma(z)$, we note that the flux is 
	inversely proportional to the beam area $w^2(z)$, as $N_\gamma(z)\propto 1/w^2(z)$. 
	By expressing $w^2(z)$ in terms of the beam area at the focus $w_0^2$, and the Rayleigh range,  we get   
	\bea
	N_\gamma (z) \propto \frac{1}{w^2(z)} \propto  \frac{1}{w_0^2} \frac{z_R^2}{z^2+z_R^2}\,, 
	\eea
	which reproduces the naive scaling of $N_\gamma (z)\propto 1/z^2$ for $z\gg z_R$. 
	The Rayleigh range can be expressed in terms of in terms of focused beam area  as
	\bea
	z_R = \frac{\pi w_0^2}{\lambda} \sim  \frac{\lambda}{ \Omega}\,,  
	\eea
	where, $\Omega$ is the angular spread in square radians, and $\lambda$ is the  wavelength. 
	For an extremely collimated SR radiation with  $\Omega\sim 10^{-10}\,{\rm sr}$, and $\lambda\sim 10^{-10}\,{\rm m}$, the Rayleigh range is $z_R\sim \mathcal{O}(\rm m)$. 
	Thus, we want to consider a set-up whose Rayleigh range is either large, or comparable to the distance of our interest.  
	\\
	
	For that purpose, we consider focusing of the emitted photon by using a lens. 
	For a lens with focus length $f$, and refractive index $n$, the effective radius of curvature, $R_{\rm eff}$, can be written as~\cite{Hecht}
	\bea
	R_{\rm eff}= (n-1) f\,.
	\label{eq:Reff}
	\eea
 The resonant nuclear contribution to the refractive index of a material can be written as~\cite{PhysRev.171.417}, 
	\bea
	\!\!\!\!\!\!\!\!
	n(E_\gamma)-1 = \frac{n_{\rm nuc}\, \sigma_{\rm res}(E_\gamma)\, f_{\rm LM}}{2\,E_\gamma}\simeq \frac{2\,n_{\rm nuc}}{E_\gamma^3}\,,
	\eea
	where, $n_{\rm nuc}\simeq (\keV)^3$ is the nuclei number density 
  in a material, and $f_{\rm LM}$ is the Lamb-\Moss factor which is replaced by its resonant value of $2/\pi$ to obtain the last equality. 
	Thus, we obtain the range over which the emitted photons would remain parallel as
	\bea
	\!\!\!\!\!\!\!\!
	z_R\simeq \frac{\pi R_{\rm eff}^2}{\lambda}\sim 3.2\,{\rm km} \left(\frac{20\keV}{E_\gamma}\right)^5 \left(\frac{f}{1\,{\rm m}}\right)^2\,.
	\eea 
	Thus the flux at the \Moss receiver would be the same as the \Moss emitter. 
	Using Eq.~\eqref{Eq:deltaE_exp} and Eq.~\eqref{eq:fulx_moss_source}, we obtain
	\bea
	\Delta E = \sqrt{\frac{\Gamma\,(\Delta E)_{\rm SR}}{(N_{\gamma})_{\rm SR}}}\equiv \sqrt{\frac{\Gamma}{\Phi_{\rm SR}}}\,,
	\label{Eq:eq_deltaE_final}
	\eea
	where the spectral flux of a synchrotron source $\Phi_{\rm SR}$ is defined as the ratio of the flux per bandwidth. 
For a given integration time, using the above equation we obtain the sensitivity of the experiment by $\Phi_{\rm SR}\to \Phi_{\rm SR} \times 
(t_{\rm int}/{\rm s})\,{\rm Min}\left[1,(\tau_{\rm coh}/t_{\rm int})^{1/2}\right]$~\footnote{Note that, in Tab.~\ref{Tab:table_eliment_list}, $(N_\gamma)_{\rm SR}$ ($\Phi_{\rm SR}$) is given in the unit of  $\gamma/{\rm s}$ ($\gamma/{\eV\cdot \rm s}$) where here it is dimensionless ($\gamma/\eV$).}. 
	\\
	
	Due to the spread in the energy domain, the emitted photons will have a longitudinal coherence length of $\sim 2\pi/\Gamma$. 
	Although the relative phase information is regained at twice the  distance, we consider a conservative approach, and
	limit the maximum achievable baseline for a given \Moss transition as 
	\bea
	L\lesssim 200\,{\rm m}\left(\frac{\tau_{\rm dec}}{100\,{\rm ns}}\right)\,.
 \label{eq:max_baseline}
	\eea

Furthermore, we require the DM field to remain coherent over the baseline. 
For a separation of $L$ between the \Moss emitter and the absorber, this 
 provides a constraint on DM mass as, 
	\bea
	m_\phi \lesssim 4.1 \times 10^{-4}\eV\left(\frac{3\,{\rm km}}{L}\right)\,.
	\label{Eq:eq_coh_time}
	\eea

\skipnew{\color{myred}
Note that for \eqref{Eqn:MainShift} to be valid,  the dark matter mass $m_{\phi} \lesssim \Gamma$ where $\Gamma$ is the decay width of the \Moss transition. This is because the decaying nucleus takes a time $\sim 1/\Gamma$ to produce the photon and modulations at frequencies higher than $\Gamma$ result in the production of sidebands around the transition energy as opposed to a homogeneous shift to the produced energy. The latter is the case of interest here and it occurs in the limit $m_{\phi} \lesssim  \Gamma$. }

Moreover, to observe a homogeneous shift in the energy levels due to oscillating DM, we require the decay time of the nucleus $\tau_{\rm dec}$, to be shorter than the DM oscillation time. 
Thus we get, 
	\bea
	\!\!\!\!\!\!\!\!
	t_{\rm osc}=\frac{2\pi}{m_\phi}\gtrsim \tau_{\rm dec} 
 \Rightarrow m_\phi\lesssim 2\pi\,\tau_{\rm dec}\,.
	\label{Eq:eq_stability}
	\eea
 For DM oscillation frequencies higher than the above limit results in inhomogeneous energy shifts, and produces sidebands around the transition. 
\skipnew{
 to observe our effect, the spread of both the emission, and the absorption lines in the time-domain, should not exceed exceed the oscillation period of DM induced $\Delta E$. Thus we require, 
	\bea
	\!\!\!\!\!\!\!\!
	t_{\rm osc}=\frac{2\pi}{m_\phi}\gtrsim \tau_{\rm dec} 
 \Rightarrow m_\phi\lesssim 2\pi\,\tau_{\rm dec}\,,
	\label{Eq:eq_stability}
	\eea
	where, $\Gamma_{\rm nat}$ is the natural line-width of the transition. 
 }
Combining Eq.~\eqref{Eq:eq_coh_time} and Eq.~\eqref{Eq:eq_stability}, we get an upper bound on the DM mass as,
	\bea
	\!\!\!\!\!\!\!\!\!\!\!\!\!\!
	m_\phi\lesssim 8.3 \times 10^{-5}\eV\times {\rm Min}\left[\frac{15\,{\rm km}}{L},\frac{50\,{\rm ps}}{\tau_{\rm dec}} \right].
	\label{Eq:eq_mphi_upper_bound}
	\eea	
	\skipnew{
	To illustrate the reach of our proposed method, we list a few transitions with decay time ranging from $\tau_{\rm dec}\sim \mathcal{O}(10\,\mu\rm{s})$ to $\mathcal{O}(50\, {\rm ps})$ and transition energy from $E_{0}\sim (5-75)\keV$ in Table~\ref{Tab:table_eliment_list}. We chose the  transitions as benchmark points, and to highlight a few different features. 
	Choosing 
	a variety of \Moss nuclei with different decay times, would enable us to probe ULDM over a wide range of frequencies.  
	As can be seen from Eq.~\eqref{Eq:eq_mphi_upper_bound}, 
 a \Moss spectroscopy with \ce{^{187}Os} 
	would allow us to probe ULDM upto a mass of $ 10^{-4}\eV$, whereas with \ce{^{181}Ta} nuclei, the reach is limited upto  $m_\phi\sim 5\times 10^{-10}\eV$. 
	On the other hand, maximum achievable baseline 
with \ce{^{181}Ta} would be $\sim 2\,\rm{km}$ as opposed to $10\,{\rm cm}$ for \ce{^{187}Os} as discussed in Eq.~\eqref{eq:max_baseline}~\footnote{ 
Note that, this is a conservative estimate, and tweaking the set-up may result in a baseline not restricted by this requirement.}. 
	On top of this, the sensitivity of the experiment scales 
    as $\Delta E\propto 1/\tau_{\rm dec}^{1/2}$ (c.f. Eq.~\eqref{Eq:eq_deltaE_final}). 
    Thus, overall with current  technology, a \Moss nuclei with larger decay time is better suited for probing DM.  } 
	
	\skipnew{for a transition is  SMS experiment with \ce{^{181}Ta} would 
		enable us to achieve a baseline of $\sim 2\,\rm{km}$, whereas 
		for \ce{^{187}Os} maximum distance between the \Moss emitter and the absorber is limited to $10\,{\rm cm}$. 
		The most common transition for \Moss spectroscopy is \ce{^{57}Fe}. 
		In Table~\ref{Tab:table_eliment_list}, we show the relevant experimental parameters to use a \ce{^{57}Fe} based SMS as a DM sensor.}

	\skipnew{
		As the \Moss transitions have linewidths of $\mathcal{O}(10^{-9}-10^{-6})\eV$, one might expect that the coherence time consideration (a la Eq.~\eqref{Eq:eq_coh_time}) provides a weaker constraint on $m_\phi$, than the stability consideration (c.f. Eq.~\eqref{Eq:eq_stability}). 
		However, depending on the transition we are considering, it could be altered which we discuss in the next section, Section~\ref{sec:tech}. 
	}

	\section{Monochromators and Beams} \label{sec:tech}

	Currently several monochromators exist all around the world; such as Advance Photon Lab (APS) in USA~\cite{apsanl}, ESRF in Europe~\cite{esrf}, and SPring-8 in Japan~\cite{SPring8}. 
	These are operating in the energy range of $\mathcal{O}(5-130)\keV$~\cite{osti_1543138,10.1063/1.2436213,Yoda2019-qj}. 
    To perform SMS of a transition, we require a  beamline whose energy range covers the transition energy of the said sample. 
	For example BLXU09 beam line at SPring-8, and 3-ID beam line at the APS are currently being used to perform SMS of various transitions 
 in the range of $(5-20)\keV$, whereas the ID18 beam line at ESRF, is being used to perform SMS of various transitions spanning energy range of $(5-70)\keV$. 
 To perform SMS with a transition whose energy is higher than $70\keV$, 11-ID beam line at the APS can be used. 
	See~\cite{Yoda2019-qj,10.1007/978-94-010-0045-1_2,Seto2017-yy,10.1063/1.2436213} and refs. therein for a comprehensive list of the elements (along with all the relevant parameters involving each transitions) 
	that can be used to perform SMS. 
	\\

	In Table~\ref{Tab:table_eliment_list}, the energy of the first two listed transitions is in the range of $\sim (5-20)\keV$. 
	Currently, SMS with \ce{^{181}Ta} and \ce{^{57}Fe} can be performed with the BLXU09 beam-line (among many others). 
 Currently, this beam line has achieved a flux of $6.3\times 10^{12}\,\gamma/{\rm s}$ with a bandwidth of $35.7\meV$~\cite{yasui2023bl09xu}. 
	The transition energy of \ce{^{187}Os} is $74.4\keV$, and thus can not be probed by the BLXU09 beam~\footnote{\ce{^{187}Os} has another transition with decay time of $3.4$ ns and $E_0\sim 10\keV$~\cite{PhysRevB.91.224102,Basunia:2009hgg}. 
		BLXU09/3-ID beam can be used to conduct  SMS of this isotope.}. 
	20-ID-D beam line at the APS~\cite{apsanl} is operating in the energy range of $35-120\,$keV, with an obtained flux of $10^{14}\,\gamma/{\rm s}$ at a resolution of $70\eV$. 
	This beam line 
	can be used to conduct SMS with \ce{^{187}Os}. 
	\skipnew{{\color{red}These specific numbers for the flux and the energy resolution at the beam lines are obtained at a specific peak energy. 
			For example at BLXU09 these numbers are obtained at the peak energy of $7.94\keV$, whereas for 20-ID-D, these are reached at $70\keV$. 
			Purpose of these quoted numbers is to demonstrate the reach of our proposed method with the assumption that the specific beam can reach similar numbers to cater for the listed transitions. }}
	\\
\skipnew{	
	A large number of beam lines are currently operating in the energy range of our interest with bandwidth as low as $10^{-8}\eV$~\cite{mitsui2022rayleigh}, and a flux as high as $4\times 10^{13}\,\gamma/{\rm s}$~\cite{apsanl,SPring8}. 
	As the sensitivity of the proposed experiment depends on the spectral flux, an improvement in either the synchrotron bandwidth or the flux (or both) 
	will vastly extend the reach of DM detection. 
	Several beam lines are already working on implementing changes to obtain higher spectral flux while maintaining a good energy resolution. 
	For example, the 3-ID beam line at the APS  is already operating in several modes achieving spectral flux in the range of $(5\times 10^{11}-2\times 10^{13}) \, \gamma/(\rm s\cdot \eV)$ ($N_\gamma=5\times 10^{9}\,\gamma/{\rm s}$ with a resolution of $10\meV$, and/or flux of $2\times 10^{13}\,\gamma/{\rm s}$ with $(\Delta E)_{\rm SR}\sim\eV$) ~\cite{apsanl}. 
    Their aim is to reach a resolution of $0.6\meV$ while maintaining a high flux~\cite{osti_1543138}. 
    This would lead to a spectral flux of $\Phi_{\rm SR} \simeq 3\times 10^{17}\, \gamma/({\rm s}\cdot \eV)$, resulting in an immediate improvement of  $\mathcal{O}(10^3)$ over the existing setup.}
	\\
	
	\skipnew{\color{myred}
		In Table~\ref{Tab:table_eliment_list}, we chose the  transitions as benchmark points, and to highlight a few different features. 
		Choosing 
		\Moss transitions with different decay time would enable us to probe ULDM over a wide-range of parameter space. 
		As already noted, and can be seen from Eq.~\eqref{Eq:eq_mphi_upper_bound}, the maximum probable DM mass is dictated by the decay time of the transition as $m_\phi \propto 1/\tau_{\rm dec}$. 
		Thus, a \Moss spectroscopy with \ce{^{187}Os} 
		would allow us to probe ULDM mass upto  $10^{-4}\eV$, whereas with a SMS of  \ce{^{181}Ta}, we can probe ULDM upto a mass of $5\times 10^{-10}\eV$. 
		On the other hand, maximum achievable baseline of a given transition is limited by its longitudinal coherence length as $L\propto\tau_{\rm dec}$. 
		Thus 
		
		for a transition is  SMS experiment with \ce{^{181}Ta} would 
		enable us to achieve a baseline of $\sim 2\,\rm{km}$, whereas 
		for \ce{^{187}Os} maximum distance between the \Moss emitter and the absorber is limited to $10\,{\rm cm}$. 
		The most common transition for \Moss spectroscopy is \ce{^{57}Fe}. 
		In Table~\ref{Tab:table_eliment_list}, we show the relevant experimental parameters to use a \ce{^{57}Fe} based SMS as a DM sensor.} 
	

	\skipnew{ 
		{\color{red}As shown in~\cite{Yaroslavtsev:gy5040}, highest intensity at a specified synchrotron source width can be obtained by varying both the angular position, and the temperature of the synchrotron set-up crystal. }}
	
	\begin{table*}[t!]
		\begin{center}
			\begin{tabular}{|p{1.9cm}|p{1.2cm}|p{1.8cm}|p{1.5cm}||p{1.8cm}|p{2.0cm}|p{2.0cm}|}
				\hline
				\multicolumn{4}{|c||}{Transition List} & \multicolumn{3}{|c|}{Beam} \\
				\hline
				\centering
				Transition & $E_{0}$ (keV)& $\Gamma_{\rm exp}$ (eV) & $\tau_{\rm dec}$  & $(\Delta E)_{\rm SR}$ & $(N_\gamma)_{\rm SR}$ $\gamma/\rm s$ & $\Phi_{\rm SR}$ $\gamma/(\rm s\cdot\eV)$  \\
				\hline
				\ce{^{181}_{73}Ta}~\cite{Wu:2005obv} & 6.2 & $5.5\times 10^{-10}$ & 8.7 $\mu$s & 35.7 \meV& $6.3\times 10^{12}$ & $1.76\times 10^{14}$\\
				\ce{^{57}_{26}Fe}~\cite{Bhat:1998ina}  & 14.4 & $5\times 10^{-9}$& 141.8 ns & 35.7 \meV& $6.3\times 10^{12}$ & $1.76\times 10^{14}$\\
				\ce{^{187}_{76}Os}~\cite{Malmskog:1971tmu,Basunia:2009hgg} & 74.4 & $1.2\times 10^{-5}$ & 0.0534 ns & 70\eV & $1\times 10^{14}$ & $1.43\times 10^{12}$ \\
				\hline
			\end{tabular}
		\end{center}
		\caption{List of relevant parameters for a few \Moss transitions that can be used to probe DM (see~\cite{Yoda2019-qj,10.1007/978-94-010-0045-1_2,Seto2017-yy,10.1063/1.2436213} and refs. therein for a more comprehensive list of the elements along with all the relevant synchrotron source  parameters involving each transitions, that can be used to perform SMS). 
	The listed transitions are chosen as benchmarks and also to highlight various facets of our proposed method.  We divide the table in two main columns: ``Transition List'' -   providing information about a specific transition, and ``Beam" - providing information about currently operating synchrotron beam that can be used to perform a SMS of that specific transition. 
Left to right: we outline a transition, its energy $E_0$, experimentally observed line width, $\Gamma_{\rm exp}$, decay time $\tau_{\rm dec}$, 
			the energy resolution $(\Delta E)_{\rm SR}$, the flux $(N_\gamma)_{\rm SR}$ per second, and the spectral flux $\Phi_{\rm SR}=(N_\gamma)_{\rm SR}/(\Delta E)_{\rm SR}$ of the synchrotron source. We quote the experimentally observed line-widths for \ce{^{181}Ta}, and \ce{^{57}Fe}, whereas for \ce{^{187}Os} 
			we obtain it from the observed decay time. 
			Note that, for \ce{^{181}Ta}, the observed line width is broadened compare to the natural one by almost a factor of $7.5$~\cite{DORNOW1979491}. In such case, one should use the observed width to estimate the sensitivity.} 
		\label{Tab:table_eliment_list}
	\end{table*}

	\skipnew{
		\color{red}In Table~\ref{Tab:table_eliment_list}, we discuss various quantities related to a \Moss transition that are of relevance for this work. 
		In the table, (left to right) we outline a transition, its energy $E_0$, experimentally observed line width, $\Gamma_{\rm exp}$, and the effective flux, $N_\gamma$, received at the \Moss source after an integration time of a month. 
		The next two columns demonstrate (left to right) the energy resolution, $(\Delta E)_{\rm SR}$, and the flux, $(N_\gamma)_{\rm SR}$ per second, of the synchrotron source that can be used to perform a SMS of that specific transition. 
		These two columns are clubbed together, and put under ``Beam". 
		We quote the experimentally observed line-widths for \ce{^{181}Ta}, and \ce{^{57}Fe}, whereas for the last two transitions we obtain it from the observed decay time. 
		Note that, for \ce{^{181}Ta}, the observed line width is broadened compare to the natural one by almost a factor of $7.5$~\cite{DORNOW1979491}.
	}

	\section{Detection Sensitivity}\label{sec:sensitivity}

	In this section, we consider different interactions of DM with the low energy SM, and calculate the nuclear  energy shift due to that interaction. 
	
	Let us consider a linear coupling between the ULDM, $\phi$, with the up ($u$), and down ($d$) quarks, gluons ($G^{a}_{\mu\nu}$), and photons ($F_{\mu\nu}$) as,
	\bea
	\!\!\!\!\!\!\!\!\!\!\!
	\mathcal{L}\supset \sum_{\psi=u,d}\!-\frac{\phi}{f_\psi} m_\psi \bar \psi \psi - \frac{\alpha\,\phi}{4\,f_\gamma} F^2-\frac{\beta(\alpha_s)}{4\alpha_s}\frac{\phi}{f_g} G^2\,,
	\label{eq:lagrangian}
	\eea
	where, $G^2=G_{a\mu\nu}G^{a\mu\nu}$ with $a$ being the color index, $F^2=F_{\mu\nu}F^{\mu\nu}$, $m_\psi$ is the mass of $\psi$, $\alpha$ is the fine structure constant,  $\beta(\alpha_s)$ is the beta function of the strong coupling constant $\alpha_s$,  and $f_i$ denotes the interaction strength with various modules. 

	\subsection{alpha modulus}

	The coupling $\frac{\alpha\,\phi}{4\,f_\gamma} F^2$ induces an oscillating component to the fine structure constant due to the time dependent $\phi$ background as given in Eq.~\eqref{eq:phi_bg}. 
	The shift in $\alpha$ can be written as
	\bea
	\!\!\!\!\!\!\!\!\!\!\!\!
	\alpha \to \alpha - \frac{\alpha^2\phi}{f_\gamma}
	\Rightarrow \frac{\Delta\alpha}{\alpha(0)}=\frac{\alpha(\phi)-\alpha(0)}{\alpha(0)} \simeq -\frac{\alpha\,\phi}{f_\gamma}.
	\eea 
	
	The dominant electromagnetic effect in the nuclear energy is encoded in the $\alpha$ dependence of the nucleon masses and nuclear binding energy~\cite{Damour:2010rp}. 
	Most importantly, due to the change in $\alpha$, the electromagnetic self energy of a nucleus of size  $r_{\rm nuc}$, with $Z$ protons and $A$ nucleons changes as $\sim Z^2 \Delta\alpha/r_{\rm nuc}$.
 \skipnew{
 would change as
	\bea
	\Delta E_{\rm bind}\simeq \frac{Z^2 \Delta\alpha }{A^{1/3}r} = \frac{Z^2 \alpha}{A^{1/3}r}\frac{\Delta\alpha}{\alpha}
	\,,
	\eea
	where, $r$ is the average position of a nucleon. 
 } 
As the \Moss transitions involve transitions between nearly degenerate energy levels, we estimate the energy due to one nucleon transition. 
Thus we obtain the energy shift due to change of $\alpha$ as~\cite{Gratta:2020hyp}, 
	\bea
	\!\!\!\!\!\!\!\!\!\!\!\!
	\Delta E \simeq \frac{Z\Delta\alpha}{r_{\rm nuc}}
 \simeq -\frac{Z\alpha^2}{A^{1/3}r_0}\frac{\Delta\phi}{f_\gamma}
	\,,
	\label{eq:delE_alpha}
	\eea
	where, we take the position of the transitioning nucleon to be the nuclear size with $r_{\rm nuc}=A^{1/3}\,r_0 $, and $r_0\sim 1.2\,{\rm fm}$ is some typical nuclear scale~\cite{krane1991introductory}. 
 

	\subsection{Yukawa modulus}
	
	The low energy dynamics of nuclear physics, can largely be described by the physics in the Isospin limit. 
	Thus, we consider $\phi$ coupling to the Isospin symmetric combination of the $u$, and $d$ quarks masses. 
	We define $\hat m = (m_u+m_d)/2$, and $2 \hat m/f_{\hat m} = (m_u/f_u + m_d/f_d)$, and obtain
	\bea
	\!\!\!\!\!\!\!\!\!\!
	\hat m\to \hat m-\frac{\hat m\,\phi}{f_{\hat m}}\Rightarrow \frac{\Delta \hat{m}}{\hat m} =\frac{\phi}{f_{\hat m}}\,,
	\eea
using Eq.~\eqref{eq:lagrangian}. 
As the light quark masses largely affect the pion mass ($m_\pi$) as $m_\pi^2\propto \hat m$, we consider the one-pion exchange (OPE) potential between two nucleons to estimate the change in nuclear energy $\Delta E$. 
	Following the notation of~\cite{Machleidt:1987hj,Machleidt:2001rw,HitoshilectureNote}, the OPE potential can be written as,
	\bea
	V_{\rm OPE}(r)=-\frac{g_{\pi NN}^2}{4\pi}\frac{m_\pi^2}{4\,m_N^2}\frac{e^{-m_\pi r}}{r}\,,
	\eea
	where, $g_{\pi NN}^2/(4\pi)=15$ is the effective pion nucleon interaction strength. 
	For a distance $r\lesssim 1/m_\pi$,  we get,
	\bea
	\Delta E &\simeq &  \frac{g_{\pi NN}^2}{(4\pi)\,4\, r} \left[\frac{\partial(m_\pi^2/m_N^2)}{\partial\ln m_\pi^2} \frac{\partial\ln m_\pi^2}{\partial\ln\hat m}\right] \frac{\Delta \hat{m}}{\hat m} \nn\\
	&\simeq& 12.3 \MeV\, \frac{\Delta\phi}{f_{\hat m}}\,,
	\label{eq:delE_mhat}
	\eea
	where, we use $\partial \ln m_\pi^2/\partial\ln\hat m=1$, $\partial\ln m_N/\partial\ln m_\pi^2\simeq 0.06$~\cite{Kim:2022ype}, and set $r=r_0$. 

	\subsection{Gluon modulus}
	
	Similar to the $\alpha$ coupling, $\frac{\beta(\alpha_s)}{4\alpha_s}\frac{\phi}{f_g} G^2$ interaction shifts the strong coupling constant, and $\phi$ induced change to $\alpha_s$ can be written as 
	\bea
	\frac{\Delta\alpha_s}{\alpha_s(0)}= \frac{\alpha_s(\phi)-\alpha_s(0)}{\alpha_s(0)} \simeq -\frac{\beta(\alpha_s)}{\alpha_s}\frac{\phi}{f_g}\,.
	\label{Eq:alphas_phi}
	\eea
	As the QCD scale $\Lambda_{\rm QCD}$, depends on $\alpha_s$, we obtain the $\phi$ dependence of $\Lambda_{\rm QCD}$ as,  
	\bea
	\frac{\partial\ln\Lambda_{\rm IR}}{\partial\phi} \simeq - \frac{\alpha_s}{\beta(\alpha_s)} \frac{\partial\ln\alpha_s}{\partial\phi} = \frac{1}{f_g}\,,
	\label{Eq:lambdaIR_phi}
	\eea
	using Eq.~\eqref{Eq:alphas_phi}. 
 To calculate the change in the nuclear energy 
due to the $\phi$ dependence of $\alpha_s$, we note that any shift in $\Lambda_{\rm QCD}$ corresponds to a change in the whole hadronic spectrum. 
A small $\phi$ induced change can be estimated by considering OPE potential with $\phi$ dependent pion-nucleon coupling constant as, 
	\bea
	\Delta E &\simeq &  \left[\frac{g_{\pi NN}^2/(4\pi)}{4\, r_0} \frac{\partial(m_\pi^2/m_N^2)}{\partial\ln \Lambda_{\rm IR}} \frac{\Delta \Lambda_{\rm IR}}{\Lambda_{\rm IR}}\right.\nn\\
	&& \left.+\,\frac{m_\pi^2}{4r_0\,m_N^2}\frac{\partial (g_{\pi NN}^2/4\pi)}{\partial\ln\alpha_s}  \frac{\Delta \alpha_s}{\alpha_s}\right]\nn\\
	&\simeq& 0.1 \GeV\, \frac{\Delta \phi}{f_g}\,,
	\label{eq:delE_gluon}
	\eea
where we use $\partial\ln (g_{\pi NN}^2/4\pi)/\partial\ln\alpha_s =1$. 
\skipnew{
we consider OPE potential with $\phi$ dependent pion-nucleon coupling constant.   

 Also, as the change of the $\Lambda_{\rm QCD}$ stems from changing the strong coupling, it also affects the hadronic  couplings as well. 
	However, for our purpose, we are interested in a simple order of magnitude estimate without the subtleties of nuclear physics near the confinement scale. 
	As, most of the nucleon mass comes from the gluonic part~\cite{Becher:1999he, Shifman:1978zn}, rough estimate suggests, $m_N\sim \alpha_s/r_0$ where, $m_N\sim 930\MeV$ is the nucleon mass at the chiral limit~\cite{Becher:1999he}. 
	Assuming the change of the meson-nucleon effective coupling is linear with that of the strong coupling constant, i.e. $\partial\ln (g_{\pi NN}^2/4\pi)/\partial\ln\alpha_s =1$, we obtain 
	\bea
	\Delta E &\simeq &  \left[\frac{g_{\pi NN}^2/(4\pi)}{4\, r_0} \frac{\partial(m_\pi^2/m_N^2)}{\partial\ln \Lambda_{\rm IR}} \frac{\Delta \Lambda_{\rm IR}}{\Lambda_{\rm IR}}\right.\nn\\
	&& \left.+\,\frac{m_\pi^2}{4r_0\,m_N^2}\frac{\partial (g_{\pi NN}^2/4\pi)}{\partial\ln\alpha_s}  \frac{\Delta \alpha_s}{\alpha_s}\right]\nn\\
	&\simeq& 0.1 \GeV\, \frac{\Delta \phi}{f_g}\,,
	\label{eq:delE_gluon}
	\eea
	from the OPE potential using Eqs.~(\ref{Eq:alphas_phi} -- \ref{Eq:lambdaIR_phi}). We have also used $2 \beta(g_s)/g_s=-9\alpha_s/(2\pi)$. } 
	\skipnew{
		\subsection{Dilaton Coupling}

		In this subsection, we consider a special case, where the scalar field, $\phi$, couples to the trace of trace of the energy-momentum tensor of the SM, $T^{\mu}_{\mu,\,\rm SM}$ as $\mathcal{L}\supset -\phi/f_{\rm dil}\, T^{\mu}_{\mu,\,\rm SM}$ with $f_{\rm dil}$ being the effective dilaton interaction strength, and 
		\bea
		\!\!\!\!\!\!\!\!\!\!\!\!\!\!\!
		T^\mu_\mu= \!\!\sum_{\psi}\!\!\! m_\psi (1-\gamma_{m_\psi}) \bar \psi \psi + \!\!\frac{\beta(\alpha)}{4\,\alpha} F^2\!+\!\frac{\beta(g_s)}{2g_s} G^2,
		\label{eq:tmumu}
		\eea
		where, summation over $\psi$ runs over all the SM fermions, but for our purpose we would only consider $u$, and $d$ quarks, $\gamma_{m_\psi}$ is the gamma function of the particular fermion $\psi$, $\beta(\alpha)=2\alpha^2/(3 \pi)$ is the beta function of the electromagnetic interaction, and $f_{\rm dil}$ is the effective dilaton coupling strength. 
		As, gravity couples to the Ricci scalar $R$ which is proportional to $T^{\mu}_{\mu,\,\rm SM}$ (using Einstein’s equation), a dilaton coupling to the SM is universal, and thus, can not be probed by the experiments looking for EP violations~\cite{Banerjee:2022sqg}. 
		However, a dilaton would mediate a Yukawa type long range force with interaction strength given by its coupling to the SM, leading to a deviation from inverse-square law force. 
		Hence, it can be probed by composition independent fifth force search experiments~\cite{Fischbach:1996eq}. 
		
		A light dilaton DM\skipnew{~\footnote{Note that, as shown in~\cite{Hubisz:2024hyz}, for a dilaton to be wavelike DM with a minimal misalignment mechanism, one would require enormous tuning of several theory parameters. However, this can be circumvented by considering a supersymmetric set-up~\cite{Banerjee:2024xxx}}} would induce a time-dependent shift to the nuclear energy levels, and thus can be probed by our proposal. 
		As the dilaton couples to all the fundamental fields, a dilaton would induce a change in all the fundamental constants, $g={\alpha, \hat m, \alpha_s}$, at the same time. Thus, we obtain the total energy shift by a dilaton as,
		\bea
		\!\!\!\!\!\Delta E &=& \sum_{i}\frac{\partial E}{\partial g_i}\partial g_i \skipnew{\simeq \sum_i \left(\frac{\partial E}{\partial \ln g_i}\frac{\partial\ln g_i}{\partial\phi}\right) \Delta\phi\,,\nn\\}
		\simeq \frac{m_\pi^2}{4r_0\,m_N^2}\frac{\partial (g_{\pi NN}^2/4\pi)}{\partial\ln\alpha_s} \frac{\Delta \alpha_s}{\alpha_s}\nn\\
		&\simeq & 0.11\GeV\,\frac{\Delta\phi}{f_{\rm dil}}\,,
		\label{eq:delE_dilaton}
		\eea
		where, we redefine the fermion mass to its pole mass, and 
		ignore the small energy correction due to the change of $\alpha$. 
	}

\section{Results and Discussion}\label{sec:reults_discussions}
	
	\begin{figure}[t!]
		\centering
		\includegraphics[width=\columnwidth]
		{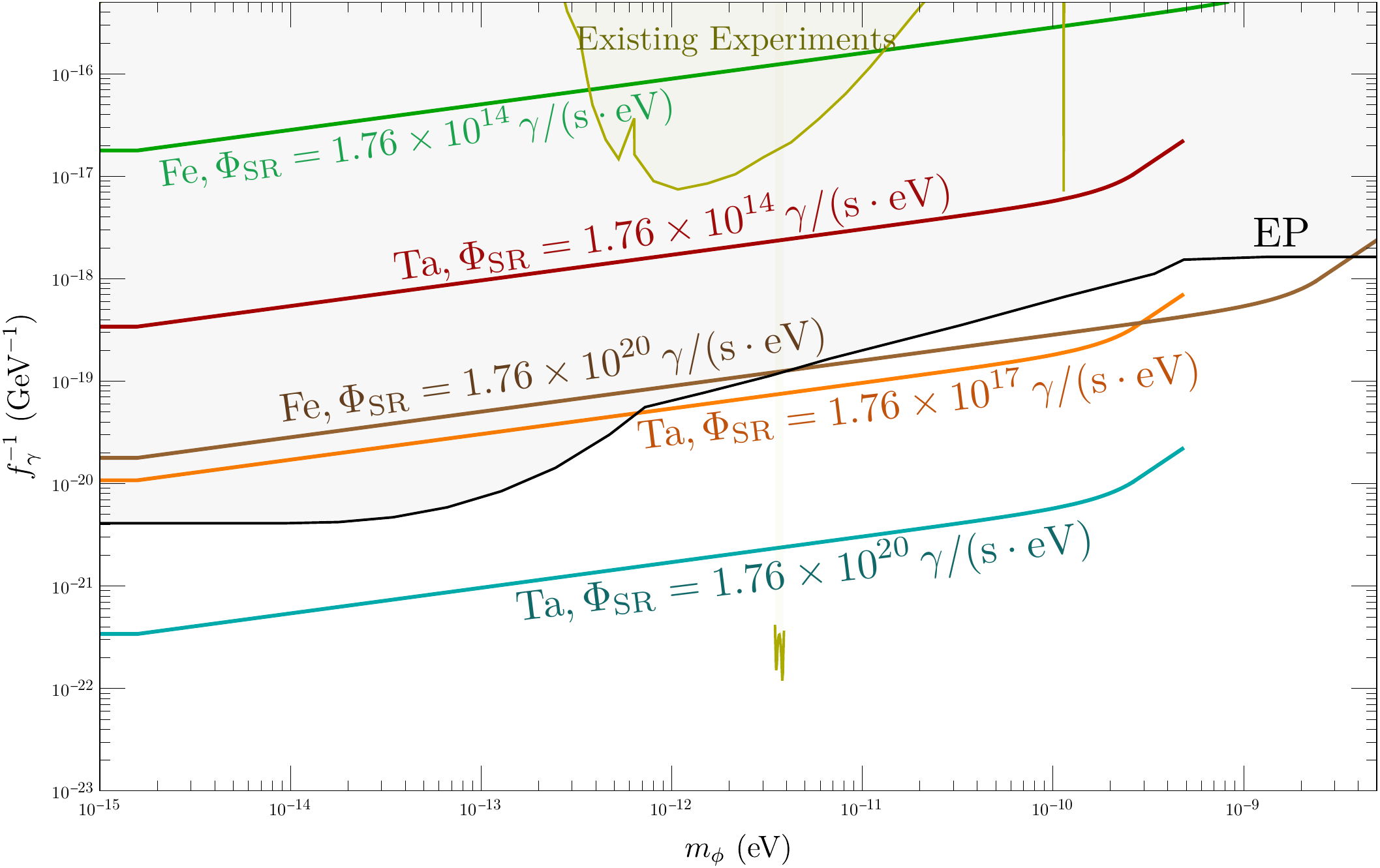}
            \centering
		\caption{Parameter space of the DM-photon interaction strength $f_\gamma^{-1}$, as a function of DM mass $m_\phi$. 
        The reach of the proposed experiment with existing synchrotron spectral flux are shown by the red (\ce{^{181}Ta}) and green (\ce{^{57}Fe}) lines. See Table~\ref{Tab:table_eliment_list} for more details. 
        Reach of a \Moss experiment with improved spectral flux with different nuclei are also depicted: orange ($10^3$ with \ce{^{181}Ta}), turquoise ($10^6$ with \ce{^{181}Ta}), and brown ($10^6$ with \ce{^{57}Fe}). 
        The projections are shown for the integration time of $1$ month, and the maximal achievable baseline. 
        The gray and the yellow shaded regions depict the bound from various equivalence principle (EP) violating tests~\cite{MICROSCOPE:2022doy,Schlamminger:2007ht,Smith:1999cr} and existing DM search experiments~\cite{PhysRevLett.126.051301,Vermeulennature2021, Antypas:2022asj} respectively.            
        }
		\label{fig:fgamma}
	\end{figure}

\begin{figure}[t!]
		\centering
		\includegraphics[width=\columnwidth]
		{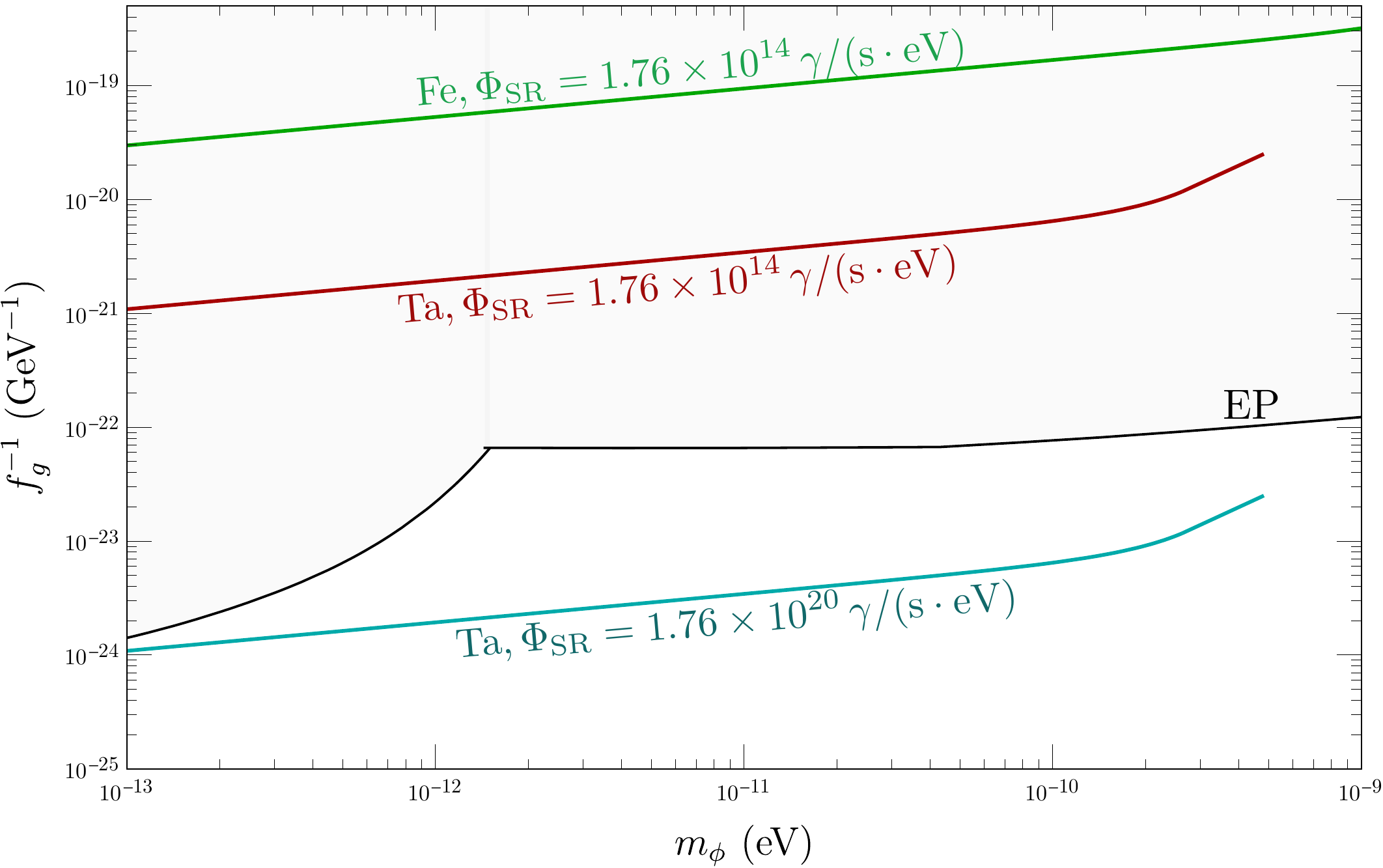}
        \includegraphics[width=\columnwidth]
		{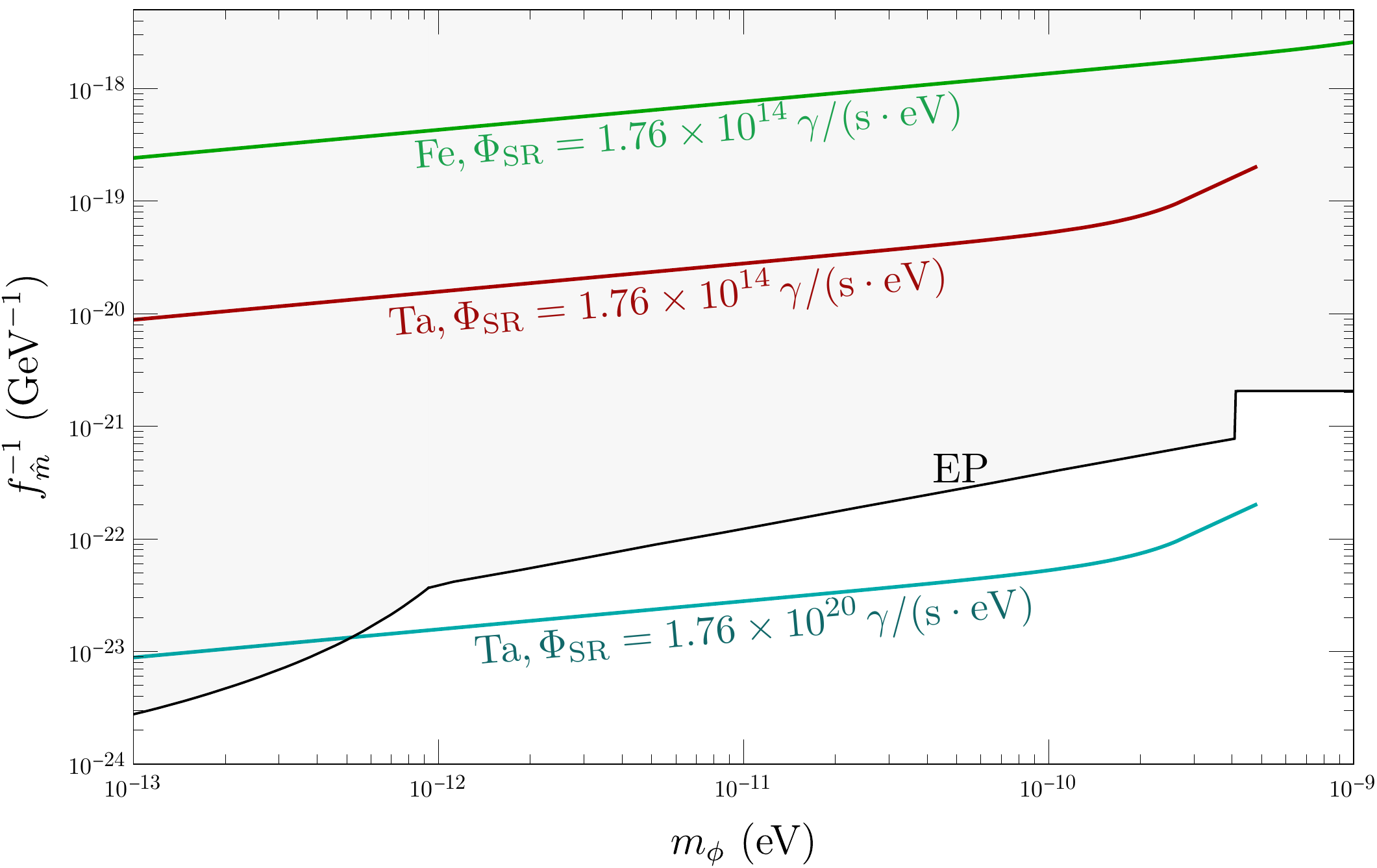}
		\caption{
			Parameter space of the DM-gluon $f_g^{-1}$ (top), and DM-quarks $f_{\hat m}^{-1}$ (bottom) interaction strength as a function of DM mass $m_\phi$. 
   Existing experimental bounds are taken from~\cite{Oswald:2021vtc}. Remaining description matches the caption of Fig.~\ref{fig:fgamma}.
		}
		\label{fig:fg}
	\end{figure}
\skipnew{
	\begin{figure}[t!]
		\centering
		\includegraphics[width=\columnwidth]
		{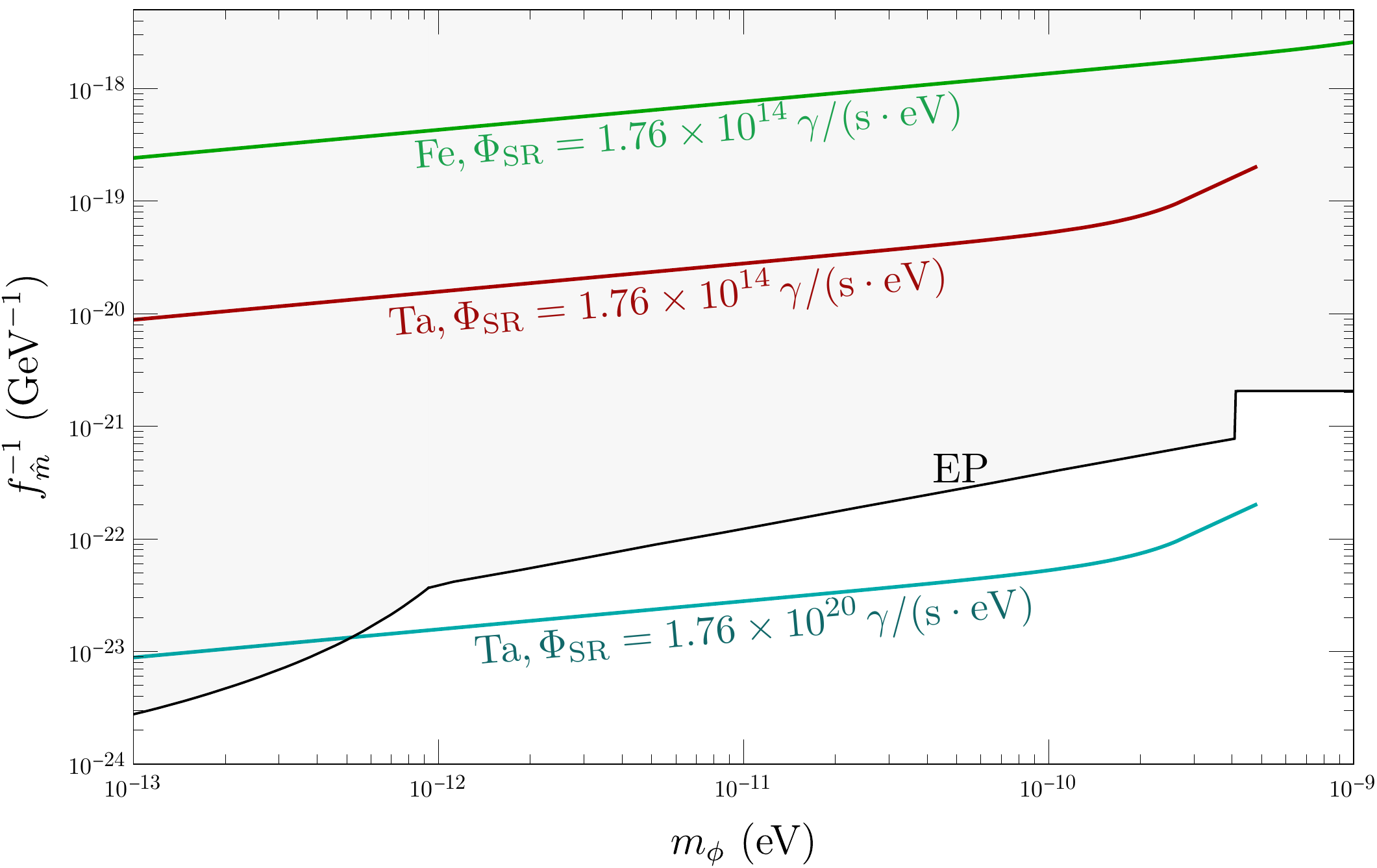}
		\caption{
			Constraints on the scalar DM-quark mass interaction strength, $f_{\hat m}^{-1}$, as a function of DM mass, $m_\phi$. 
			Existing experimental bound is taken from~\cite{Oswald:2021vtc}. Remaining description matches the caption of Fig.~\ref{fig:fgamma}.  
		}
		\label{fig:fm}
	\end{figure}
		\begin{figure}[t!]
			\centering
			\includegraphics[width=0.96\columnwidth]
			{Figures/fdil_100.pdf}
			\!\!\!\!\includegraphics[width=0.94\columnwidth]{Figures/fdil_3000.pdf}
			\caption{
				Constraints on the dilaton-SM interaction strength, $f_{\rm dil}^{-1}$, as a function of DM mass, $m_\phi$. 
				Existing experimental bounds are from~\cite{PhysRevLett.125.201302, PhysRevLett.129.241301}. The black line depicts the bound from various composition independent fifth force searches~\cite{Fischbach:1996eq}. Note that, as discussed in the main text, EP violation search experiments are insensitive to a dilaton. 
			}
			\label{fig:fdil}
		\end{figure}
	}

To illustrate the reach of our proposed method, we list a few transitions with decay time ranging from $\tau_{\rm dec}\sim \mathcal{O}(10\,\mu\rm{s})$ to $\mathcal{O}(50\, {\rm ps})$ and transition energy from $E_{0}\sim (5-75)\keV$ in Table~\ref{Tab:table_eliment_list}. 
Using the listed transitions 
as benchmarks, we show the parameter space of various dark matter-SM interaction in Fig.~\ref{fig:fgamma} and Fig.~\ref{fig:fg}. 
We plot the parameter space for each module by equating the DM induced the energy shifts, i.e Eqs.~(\ref{eq:delE_mhat}, \ref{eq:delE_alpha}, \ref{eq:delE_gluon}) with Eq.~\eqref{Eq:eq_deltaE_final} for 
an integration time of $1$ month. 
Choosing a variety of \Moss nuclei with different decay times enables us to probe ULDM over a wide range of frequencies. 
As can be seen from Eq.~\eqref{Eq:eq_mphi_upper_bound}, a \Moss spectroscopy with \ce{^{187}Os} would allow us to probe ULDM mass upto $\sim 10\,{\rm GHz}$, whereas with \ce{^{181}Ta} and \ce{^{57}Fe} nuclei, the reach is limited upto $\sim 1$ MHz. 
On the other hand, maximum achievable baseline with \ce{^{181}Ta} is $\sim 2\,\rm{km}$ as opposed to a $10\,{\rm cm}$ for \ce{^{187}Os} as can be seen from Eq.~\eqref{eq:max_baseline}~\footnote{ 
Note that, this is a conservative estimate, and tweaking the set-up may result in a baseline not restricted by this requirement.}. On top of this, the sensitivity of the experiment scales 
    as $\Delta E\propto 1/\tau_{\rm dec}^{1/2}$ (c.f. Eq.~\eqref{Eq:eq_deltaE_final}). 
    Thus, overall with current  technology, a \Moss nuclei with larger decay time is better suited for probing DM.
\\

\skipnew{
sensitivity proposed synchrotron based \Moss spectroscopy in probing ULDM which interacts linearly with the SM are shown in Fig.~\ref{fig:fgamma} and Fig.~\ref{fig:fg}.

we plot the parameter space for each module, by equating the DM induced the energy shifts, i.e Eqs.~(\ref{eq:delE_mhat},\ref{eq:delE_alpha},\ref{eq:delE_gluon}) with Eq.~\eqref{Eq:eq_deltaE_final}.
We chose the  transitions as benchmark points, and to highlight a few different features. 
}

In Fig.~\ref{fig:fgamma} and Fig.~\ref{fig:fg},  
the reach of a synchrotron based \Moss experiment with \ce{^{181} Ta} and \ce{^{57} Fe} nuclei with currently available spectral flux of $\Phi_{\rm SR} = 1.76\times 10^{14}\,\gamma/(\rm s\cdot\eV)$ are shown by the red and the green lines respectively. 
The orange, turquoise and brown lines depict the reach a \Moss experiment with improved synchrotron beam lines. 
The first two projections are made 
with \ce{^{181} Ta} with an improved spectral flux of $10^3$ and $10^6$ respectively, whereas for the brown lines, we consider  \ce{^{57} Fe} with $10^6$ improvement. 
All the projections are shown with the maximum achievable baseline for the given transition. 
Constraints from the existing ULDM searches are shown in yellow shaded region~\cite{Antypas:2022asj}, whereas the solid black lines represent the bound from various equivalence principle (EP) violation searches~\cite{MICROSCOPE:2022doy,Schlamminger:2007ht,Wagner:2012ui,Smith:1999cr}. 
\\

\skipnew{
The baseline is taken to be the allowed 
 the longitudinal coherence length of the transition. 
 The projections are shown with an integration time of $1$ month.

of $10^3$ and $10^6$  
The reach of a SMS with the most popular \Moss nuclei, \ce{^{57} Fe}, is shown by the green (current set up), and the brown ($\times\, 10^6$ improved set up) lines. 
 The baseline is taken to be the allowed 
 the longitudinal coherence length of the transition. 
 The projections are shown with an integration time of $1$ month. 
Constraints from a variety of existing ULDM searches are shown in dark yellow color~\cite{Antypas:2022asj}, whereas the solid black lines represent the bound from various equivalence principle (EP) violation searches~\cite{MICROSCOPE:2022doy,Schlamminger:2007ht,Wagner:2012ui,Smith:1999cr}.
}

To understand the projections, 
in the low mass regime, i.e. for $m_\phi\lesssim 1/L$, the sensitivity of the experiment remains independent of DM mass for a phase coherent signal. 
Dictated by the integration time, we obtain maximum experimental sensitivity for $m_\phi \lesssim  2\pi/(t_{\rm int}\beta^2) \simeq 2\times 10^{-15}\eV$.
Beyond the coherence time, the sensitivity decreases as $m_\phi^{1/4}$ till $m_\phi\sim 1/L$, and in the high mass regime, it scales as $m_\phi^{5/4}$. 
These scaling behaviours can be seen from Fig.~\ref{fig:fgamma} and Fig.~\ref{fig:fg}. 
\\
\skipnew{
\color{red}
In Fig.~(\ref{fig:fgamma},\ref{fig:fm},\ref{fig:fg}), 
the constraints from the existing ULDM searches are shown in dark yellow color~\cite{Antypas:2022asj}. 
whereas the solid black lines depict the bound from various equivalence principle (EP) violation searches~\cite{MICROSCOPE:2022doy,Schlamminger:2007ht,Wagner:2012ui,Smith:1999cr}. 
	As can be seen from Fig.~(\ref{fig:fgamma}--\ref{fig:fg}), EP violation searches provide the most stringent existing bounds on the DM parameter space.
\\
}

The reach of the proposed experiment with existing synchrotron beam lines is at par with the bounds coming from the EP violation searches. 
For the scalar-photon interaction strength the obtained reach is within a factor of $\mathcal{O}(10)$ of the EP limits, whereas for $f_g$ and $f_{\hat m}$, the obtained bounds are weaker by $\mathcal{O}(10^2)$. 
As the experimental sensitivity is $\propto \Phi_{\rm SR}^{1/2}$, an improvement of $\mathcal{O}(10^2)$ ($\mathcal{O}(10^4)$) in spectral flux would enable us to probe hitherto uncharted territory for $f_\gamma$ ($f_g$ and $f_{\hat m}$). 
Nonetheless, even with existing set up, a SMS with \ce{^181{\rm Ta}} would provide the best bound on DM-SM interaction strength; by improving the existing limits coming exclusively from experiments searching for time oscillating signals by several orders of magnitude.
\\

A large number of beam lines are currently operating with bandwidth as low as $10^{-8}\eV$~\cite{mitsui2022rayleigh}, and a flux as high as $4\times 10^{13}\,\gamma/{\rm s}$~\cite{apsanl,SPring8}. 
	As the experimental reach 
 depends on the spectral flux, an improvement in either the synchrotron bandwidth or the flux (or both) 
	will vastly extend the reach of DM detection. 
	Several beam lines are already working on implementing changes to obtain higher spectral flux while maintaining a good energy resolution. 
	For example, the 3-ID beam line at the APS  is already operating in several modes achieving spectral flux in the range of $(5\times 10^{11}-2\times 10^{13}) \, \gamma/(\rm s\cdot \eV)$ ($N_\gamma=5\times 10^{9}\,\gamma/{\rm s}$ with a resolution of $10\meV$, and/or flux of $2\times 10^{13}\,\gamma/{\rm s}$ with $(\Delta E)_{\rm SR}\sim\eV$) ~\cite{apsanl}. 
    Their aim is to reach a resolution of $0.6\meV$ while maintaining a high flux~\cite{osti_1543138}. 
    This would lead to a spectral flux of $\Phi_{\rm SR} \simeq 3\times 10^{17}\, \gamma/({\rm s}\cdot \eV)$, resulting in an immediate improvement of $\mathcal{O}(10^3)$ over the existing setup.
\\

 With an improved spectral flux of $\mathcal{O}(10^3)$, a \ce{^{181} Ta} based \Moss experiment would  enable us to probe DM-photon interaction in the $100\, {\rm Hz}$ - $100$ kHz frequency range beyond the EP limits. 
It would extend 
the DM search by a factor of $\mathcal{O}(10)$ in the coupling strength as shown by the orange line in Fig.~\ref{fig:fgamma}. 
As a target for the synchrotron facilities, in Fig.~\ref{fig:fgamma} and~\ref{fig:fg}, we show the experimental reach with an improved spectral flux of $\mathcal{O}(10^{6})$ by the turquoise and the green lines for \ce{^181{\rm Ta}} and \ce{^57{\rm Fe}} respectively. 
This would further enable us to explore the DM parameter space upto MHz frequency with DM-photon coupling strength as low as $f_\gamma\sim 3\times 10^{22}\GeV$. Thus the reach of DM search would be extended by more than two orders of magnitude beyond the current limits.  
For DM-gluon coupling, a \Moss experiment with \ce{^{181}Ta} would be able to probe previously unexcluded part of the parameter space in the mass range of $10^{-21}\eV\lesssim m_\phi\lesssim 5\times 10^{-10}\eV$. 
It would extend the DM search upon improving the existing EP limits on $f_{g}^{-1}$ by two orders of magnitude. 
For the Yukawa coupling, the improvement over EP limits is limited to  $5\times 10^{-13}\eV\lesssim m_\phi\lesssim 5\times 10^{-10}\eV$ and a factor of $5$ on $f_{\hat m}$.
As previously discussed, an improved \Moss experiment would yield the best bound on DM-photon coupling in the kHz-MHz frequency range, even when compared to other proposed experiments~\cite{Antypas:2022asj}. 
For the DM coupling with the quarks and gluons, no currently proposed experiments can surpass the EP limits for frequencies above $100$ Hz, despite the recent insight from the $8.3\eV$  transition of \ce{^{229}Th}~\cite{Fuchs:2024edo}. 
This highlights the complementarity of the proposed experiment to conventional probes.

\skipnew{
\color{red}
improve the reach in the coupling by $\mathcal{O}(10^2)$.

xtend the DM

For the DM-photon interaction,

nable us to search for DM-photon interaction as small as  with an interaction s

to improve the reach in the coupling by $\mathcal{O}(10^2)$. 

be able to probe a coupling

a would enable us to probe DM beyond the EP limits for $10^{-21}\eV\lesssim m_\phi\lesssim 3\times 10^{-9}\eV$. 
An experiment with \ce{^181{\rm Ta}} would be able 

For the gluon and the Yukawa coupling, a SMS with current set-up provides a $\mathcal{O}(10^2)$ weaker bounds than the EP experiments. 
Thus an improvement of $\mathcal{O}(10^{4})$ in spectral flux, is required to explore new regions in the DM parameter space. 
As a target, we show the reach of a \ce{^181{\rm Ta}} based \Moss experiment with $\mathcal{O}(10^{6})$ improved spectral flux by turquoise lines in Fig.~\ref{fig:fg} and ~\ref{fig:fm}. 
For DM-gluon coupling  $f_{g}^{-1}$, a \Moss experiment with \ce{^{181}Ta} would be able to probe previously unexcluded part of the parameter space in the mass range of $10^{-21}\eV\lesssim m_\phi\lesssim 5\times 10^{-10}\eV$ with an improved beam line. 
It would set by best bound on $f_{g}^{-1}$ upon improving the existing EP limits by $\mathcal{O}(10^2)$. 
For the Yukawa coupling, the improvement over EP limits is limited to  $5\times 10^{-13}\eV\lesssim m_\phi\lesssim 5\times 10^{-10}\eV$ and a factor of $5$ in coupling. These results are shown by the turquoise lines in Fig.~\ref{fig:fg} and Fig.~\ref{fig:fm} respectively. 

nucleus   based \Moss experiment  improved spectral flux by turquoise lines

e also show the experimental reach further improvements. 
$\mathcal{O}(10^{6})\times$ current flux 

As a target, we show the reach of a \ce{^181{\rm Ta}} based \Moss experiment with $\mathcal{O}(10^{6})$ improved spectral flux by turquoise lines in Fig.~\ref{fig:fg} and ~\ref{fig:fm}.

benchmark, we show depict the  further improvement would extend the dark matter search. 
For example with a spectral flux improvement of $\mathcal{O}(10^6)$, a would enable us to probe DM beyond the EP limits for $10^{-21}\eV\lesssim m_\phi\lesssim 3\times 10^{-9}\eV$. 
An experiment with \ce{^181{\rm Ta}} would be able to improve the reach in the coupling by $\mathcal{O}(10^2)$.

hitherto uncharted territory in the mass range of $10^{-11}\eV\lesssim m_\phi\lesssim 4\times 10^{-10}\eV$. 

an immideate In Fig.~\ref{fig:fgamma} and Fig.~\ref{fig:fg}, we depict the reach of the set up with an improved spectral flux of $\mathcal{O}(10^3)$ and $\mathcal{O}(10^6)$.

onsider two 

{\color{red}
As can be seen from Fig.~\ref{fig:fgamma} and Fig.~\ref{fig:fg}, with currently available synchrotron beam lines, a \ce{^181{\rm Ta}} based \Moss experiment would provide the best bound on DM-SM interaction strength.  It would improve the existing bounds coming from the experiments exclusively searching for the time oscillating signals by several orders of magnitude. 
However EP violation searches already exclude these parts of the parameter space.  
\\


bound on scalar-photon interaction strength, obtained from \ce{^181{\rm Ta}} based \Moss experiment with currently available spectral flux, is within $\mathcal{O}(10)$ of the limit from EP violation searches.

Thus an improvement of $\mathcal{O}(10^2)$ in the spectral flux, would enable us to probe hitherto uncharted territory in the mass range of $10^{-11}\eV\lesssim m_\phi\lesssim 4\times 10^{-10}\eV$. 
This projection is shown by the orange line. 
A further improvement of the synchrotron setup would vastly extend the dark matter search. 
For example with a spectral flux improvement of $\mathcal{O}(10^6)$, a would enable us to probe DM beyond the EP limits for $10^{-21}\eV\lesssim m_\phi\lesssim 3\times 10^{-9}\eV$. 
An experiment with \ce{^181{\rm Ta}} would be able to improve the reach in the coupling by $\mathcal{O}(10^2)$. 

For the gluon and the Yukawa coupling, a SMS with current set-up provides a $\mathcal{O}(10^2)$ weaker bounds than the EP experiments. 
Thus an improvement of $\mathcal{O}(10^{4})$ in spectral flux, is required to explore new regions in the DM parameter space. 
As a target, we show the reach of a \ce{^181{\rm Ta}} based \Moss experiment with $\mathcal{O}(10^{6})$ improved spectral flux by turquoise lines in Fig.~\ref{fig:fg} and ~\ref{fig:fm}. 
For DM-gluon coupling  $f_{g}^{-1}$, a \Moss experiment with \ce{^{181}Ta} would be able to probe previously unexcluded part of the parameter space in the mass range of $10^{-21}\eV\lesssim m_\phi\lesssim 5\times 10^{-10}\eV$ with an improved beam line. 
It would set by best bound on $f_{g}^{-1}$ upon improving the existing EP limits by $\mathcal{O}(10^2)$. 
For the Yukawa coupling, the improvement over EP limits is limited to  $5\times 10^{-13}\eV\lesssim m_\phi\lesssim 5\times 10^{-10}\eV$ and a factor of $5$ in coupling. These results are shown by the turquoise lines in Fig.~\ref{fig:fg} and Fig.~\ref{fig:fm} respectively. 
}
\skipnew{
For the the light quark masses, $f_{\hat m}^{-1}$, a \Moss experiment would be able to probe previously unexcluded part of the parameter space in the mass range of 
$5\times 10^{-13}\eV\lesssim m_\phi\lesssim 5\times 10^{-10}\eV$ using \ce{^{181}Ta} transition with an improved beam line.

using \ce{^{181}Ta}

a \Moss experiment will be  only 

gluonic coupling, this can be extended to even lower masses $10^{-21}\eV\lesssim m_\phi\lesssim 5\times 10^{-10}\eV$, with an improvement over EP limits by $\mathcal{O}(10^2)$ in the coupling.

both \ce{^{57}Fe} and \ce{^{181}Ta} transitions would be able to probe parameter space beyond the EP limits in the mass range of $10^{-12}\eV\lesssim m_\phi\lesssim 5\times 10^{-10}\eV$ excluding coupling of $f_g\gtrsim 5\times 10^{22}\GeV$. 

by a factor of . 

provides probing DM interaction with the gluons, 
probing the gluon and the Yukawa module, a \ce{^181{\rm Ta}} based SMS with current set-up provides
For ,  As shown in Fig.(~\ref{fig:fg} and ~\ref{fig:fm}),  a bound on the DM paremeter space which is weaker than the EP limit by a factor of $\mathcal{O}(10^2)$.

Reach of our proposed method depends on the spectral flux of the synchrotron source as $\propto \sqrt{\Phi_{\rm SR}}$ (c.f. Eq.~\eqref{Eq:eq_deltaE_final}). 
In all the figures, the bounds from a \Moss experiment with \ce{^181{\rm Ta}} and \ce{^181{\rm Fe}} with currently available spectral flux of  $\Phi_{\rm SR} = 1.76\times 10^{14}\,\gamma/(\rm s\cdot\eV)$ are shown by light red, and the dark green lines respectively.  
A \Moss experiment with \ce{^181{\rm Ta}} would provide the best bound on DM-SM interaction strength that come exclusively from DM searches. 
However EP violation searches already exclude these parts of the parameter space.  
\\

As shown in Fig.~\ref{fig:fgamma}, bound on scalar-photon interaction strength, obtained from \ce{^181{\rm Ta}} based \Moss experiment with currently available spectral flux, is within $\mathcal{O}(10)$ of the limit from EP violation searches. 
As shown in Eq.~\eqref{Eq:eq_deltaE_final}), experimental reach is proportional to the the square root of the spectral flux.
Thus with an improved beamline wi, 

An improved beamline would enable us to probe hiertho uncharted terriory in the mass range of $10^{-11}\eV\lesssim m_\phi\lesssim 4\times 10^{-10}\eV$. This is shown by the pink line in Fig~\ref{fig:fgamma}.

the an improvment of $\mathcal{O}(10^4)$

A further improvement of spectral flux

of $\mathcal{O}(10^2)$ 

of the spectral flux  
(c.f. Eq.~\eqref{Eq:eq_deltaE_final}), an improvement of $\mathcal{O}(10^2)$ of $\Phi_{\rm SR}$  would enable us to probe parameter space beyond the EP limit.

previously  parameter space   

the square root of the spectral flux, of the synchrotron source as $\propto \sqrt{\Phi_{\rm SR}}$ (c.f. Eq.~\eqref{Eq:eq_deltaE_final}), an improvement

{\color{red}
In Fig.The bound on scalar-photon interaction strength, obtained using a \Moss experiment with of \ce{^181{\rm Ta}} with existing spectral flux, is within $\mathcal{O}(10)$ of the limit from the EP violation searches.

With current spectral flux of $\Phi_{\rm SR} = 1.76\times 10^{14}\,\gamma/(\rm s\cdot\eV)$, a \Moss experiment with \ce{^{181}Ta} would provide the best existing   are shown by light red lines. dark red lines, whereas the light  

based with \ce{^{57}Fe} SMS with \ce{^{57}Fe} with currently available spectral flux $\Phi_{\rm SR} = $

We depict the bounds using an improved  Thus an improvement of

From Fig.~\ref{fig:fgamma}, we see that for \ce{^181{\rm Ta}} changes as

$t_{\rm int}\lesssim \tau_{n}$ and . 

and ,
 
Thus, for an integration time of $1$ month (for a given $L$), the maximum experimental sensitivity is achieved till $m_\phi\lesssim 2\times 10^{-15} \eV$.

given integration time and the  baseline of the experiment, the sensitivity of the experiment remains independent of DM mass in the low mass regime,  $m_\phi\lesssim 2\pi 10^6/t_{\rm int}$. 
For DM mass between $2\pi/(\beta^2t_{\rm int}) \lesssim m_\phi\lesssim 2/L$, it decreases as $m_\phi^{1/4}$, and for $m_\phi\gtrsim 2/L$, it sufferers an additional $1/m_\phi$  suppression.  and scales as $m_\phi^{5/4}$. 

for  which decreases as 

DM mass $\lesssim 1/L$ and a phase coherent signal, the sensitivity of the experiment is independent of the DM mass. 
Beyond the coherence time, the sensitivity decreases as $m_\phi^{1/4}$ till $m_\phi\sim 1/L$, and beyond that, it scales as $m_\phi^{5/4}$.  
of a SMS with \ce{^181{\rm Ta}} changes as $1/m_\phi^{1/4}$ for DM mass $m_\phi\lesssim $

 Beyond the coherence time, the sensitivity of a SMS with \ce{^181{\rm Ta}} changes as $1/m_\phi^{1/4}$ for DM mass $m_\phi\lesssim $

 , which changes to 

To understand the sensitivity of the experiment


We show the bounds with currently available spectral flux of $\Phi_{\rm SR}= 1.76\times 10^{14}$, and with an improved  
 
 As already discussed, a SMS with \ce{^{187}Os} would yield weaker bound than that of \ce{^{181}Ta} and/or \ce{^{57}Fe}, and thus are not shown here. 
 Although a SMS with \ce{^{187}Os} would enable us to probe ULDM

 The bounds are shown for various spectral flux;

 with currently available spectral flux of $\Phi_{\rm SR} = 1.76\times 10^{14}\,\gamma/(\rm s\cdot\eV)$

We use \ce{^{181}Ta} and/or \ce{^{57}Fe} transitions listed in Table~\ref{Tab:table_eliment_list} as benchmarks for the plots.   
	In all the figures, dark green lines depict the bound from a SMS with \ce{^{57}Fe} with currently available spectral flux $\Phi_{\rm SR} = $

 whereas dark red lines show the bounds for  \ce{^{181}Ta}. 
	 
	For each transition, we consider the maximum achievable baseline which is set by its longitudinal coherence length. 
	We take integration time of $1$ month. 
	
		We show the bounds from a  for Using the transitions listed in Table~\ref{Tab:table_eliment_list} as benchmarks, we plot the parameter space for each module, by equating the DM induced the energy shifts, i.e Eqs.~(\ref{eq:delE_mhat},\ref{eq:delE_alpha},\ref{eq:delE_gluon},\ref{eq:delE_dilaton}) with Eq.~\eqref{Eq:eq_deltaE_final}. The results are shown  
		We also consider two values for the distance between the emitter, and the receiver; $L=100\,{\rm m}$, and $L=3\,{\rm km}$. 
		In all the figures, we depict the projected bound coming from a \Moss experiment with \ce{^{181}Ta} in dark red, with \ce{^{57}Fe} in dark green. 
		Whereas a \Moss experiment of \ce{^{187}Os} transition having a decay time of ${\rm ns}$ is shown in purple, and the higher energy transition with $50\,{\rm ps}$ decay time is show in magenta. 
		The solid lines depict the bound from each transition with a flux noted in the $N_\gamma$ column of Table~\ref{Tab:table_eliment_list}, whereas dashed lines represent bound with a flux that can be obtained by improved synchrotron setup. }

An improvemnet of

As shown in  Fig.~\ref{fig:fgamma}, a \Moss spectroscopy of \ce{^181{\rm Ta}} with existing spectral flux would provide

 $\Phi_{\rm SR}= 1.76\times 10^{14}$ would provide a  
 
 As shown in Fig.~\ref{fig:fgamma}, bound on scalar-photon interaction strength, $f_\gamma^{-1}$, obtained using the existing spectral flux, is within $\mathcal{O}(10)$ of that of the limit from EP violation searches. 
	More interestingly, a distance of $3\, {\rm km}$ between them with existing beam line set up would enable us to probe the parameter space beyond the fifth force limits in the mass range of $10^{-11}\eV\lesssim m_\phi\lesssim 5\times 10^{-10}\eV$. 
	With an improved synchrotron source, a \Moss experiment with \ce{^{57}Fe}, would probe uncharted region of the parameter space for three orders of magnitude in mass, $3\times 10^{-12}\eV\lesssim m_\phi\lesssim 2\times 10^{-9}\eV$, whereas with \ce{^{181}Ta}, the bound improves by an order of magnitude both in mass and coupling. 
	For the DM coupling with the Yukawa module, $f_{\hat m}^{-1}$, we can only probe previously unexcluded part of the parameter space in the mass range of $2\times 10^{-11}\eV\lesssim m_\phi\lesssim 5\times 10^{-9}\eV$ using \ce{^{181}Ta} transition with an improved beam line. 
	For the gluonic coupling, both \ce{^{57}Fe} and \ce{^{181}Ta} transitions would be able to probe parameter space beyond the EP limits in the mass range of $10^{-12}\eV\lesssim m_\phi\lesssim 5\times 10^{-10}\eV$ excluding coupling of $f_g\gtrsim 5\times 10^{22}\GeV$. 
	These results are shown by dashed green, and dashed red lines in Fig.~\ref{fig:fgamma}, Fig.~\ref{fig:fm} and Fig.~\ref{fig:fg} for \ce{^{57}Fe} and \ce{^{181}Ta} respectively.

	Synchrotron based \Moss spectroscopy of \ce{^{57}Fe} and \ce{^{181}Ta} with existing beam lines (given in Tab) 
	
	and a distance of $100\,{\rm m}$ between the emitter and the absorber, would provide the best bound on the DM-SM interaction strengths that comes from exclusively dark matter searches. However a huge fraction of this parameter space is ruled out by the fifth force search experiments. 
	This results can be seen in Fig.~(\ref{fig:fgamma}--\ref{fig:fdil}) with label $L=100\,{\rm m}$. 
	As the maximum probable mass is restricted by the decay time of the excited state (c.f. Eq.~\eqref{Eq:eq_mphi_upper_bound}), 
	a \Moss spectroscopy of \ce{^{187}Os} transition with $50\,{\rm ps}$, enables us to efficiently probe DM up to mass $\mathcal{O}(10^{-4}\eV)$. 
}

}

	\section*{Conclusion}
	
Motivated by the lack of current and future experimental efforts, in this work, we propose the possibility of probing ultra-light scalar dark matter (DM) by synchrotron radiation based \Moss spectroscopy in the kHz-MHz mass range. 
Despite their exquisite sensitivity, any  frequency comparison tests of ultra-light scalar DM, has extremely limited reach in the high frequency region. 
Thus our proposed experiment is complementary to the conventional  probes. 
\\

We have analysed the reach of the proposed method in probing DM interaction with various standard model (SM) fields. 
We have shown that with existing synchrotron beams, the reach of a \Moss experiment is at per with the most stringent limits from various equivalence principle (EP) violation searches. 
An improvement of the set up would enable us to probe hitherto uncharted territory of the dark matter parameter space; by extending the DM search by several orders of magnitude beyond the EP limits.  
\\

\Moss spectroscopy is a well-established technique that plays a crucial role in the fields of chemistry and biology. 
With the advent of powerful synchrotron sources, synchrotron radiation-based \Moss spectroscopy has become increasingly significant. 
Moreover, existing synchrotron sources are continuously improving their facilities, regardless of their application in fundamental physics. 
In this work, we demonstrate that \Moss spectroscopy can serve as an excellent sensor for dark matter, especially by extending the DM search to the high-frequency region. 
Thus, this technique complements conventional atomic and nuclear setups. Furthermore, this work could be used as a physics target to enhance synchrotron facilities which in turn would further extend its sensitivity as a DM sensor. 




\skipnew{\color{red}
investigate the possibility of using \Moss spectroscopy to probe (ultra)-light dark matter (ULDM). 
The oscillating background of ULDM induces a relative shift between the energy levels of the \Moss emitter and the absorber, leading to a modulation in the absorption cross section of the emitted photons.  
The DM induced effect depends on the distance between the \Moss emitter and the absorber, thus a large baseline is needed for better efficiency. 
Also, the narrow \Moss transitions allow this setup to probe frequencies higher than conventional setups. 
Thus we propose to use the synchrotron based \Moss spectroscopy to probe ULDM in the Hz-kHz mass range.

We have analysed the reach of our proposed method in probing DM interaction with various SM fields. 
We have shown that, with existing technology, bounds from a \Moss experiment with \ce{^{181}Ta} nuclei is at per with the most stringent limits from various equivalence principle (EP) violation searches. 
An improvement of the synchrotron set up would enable us to probe hitherto uncharted territory of the dark matter parameter space. 

\Moss spectroscopy technique is known for years, and plays a crucial role in the field of chemistry and biology. 
Moreover the existing synchrotron sources are improving their facilities irrespective of its application in fundamental physics. 
In this work, we show that \Moss spectroscopy can be a great sensor for dark matter, especially extending the DM search to the high ($\gtrsim {\rm kHz}$) frequency region.  
Thus this technique is complementary  to the conventional atomic/nuclear setups. 

Moreover this work may be used as a physics target to improve the synchrotron facilities which are useful to the field of biology/chemistry irrespective of fundamental physics.    
}

\skipnew{
here are several improvements by increasing the distance upto 3 km we

, 3 km and resolution improvement we do better

bound coming from dark matter searches, 
would provide the best bound coming from dark matter searches 

shown that with with existing technology,  nuclei provides the best bound on  

the best bound coming from dark matter searches
coming exclusively from searches involving time modulating signals. 
In particular    

with existing technology, 100 m, we get the best bound coming from dark matter searches, 

albeit the region being excluded by EP violation tests (which do not rely on dark matter)

possible with conventional atomic setups and the
system is directly sensitive to physics that changes
nuclear properties. 

To maximize the sensitivity of
this setup, it is desirable to have as large a baseline
as possible between the emitter and the absorber.

enabling us to efficiently probe the DM upto mass $\mathcal{O}(10^{-4}\eV)$.

	\Moss spectroscopy despite being known for 

  {\color{mygreen}

		\\
		
		there are several improvements by increasing the distance upto 3 km we probe uncharted territory, 3 km and resolution improvement we do better than EP by orders of magnitude
	}
}

\section*{Acknowledgement}

The author would like to thank Surjeet Rajendran for collaboration in the early phase of the project. His critical comments and discussions were extremely useful in shaping the work.
The author also thanks Ercan Alp for useful comments  regarding the longitudinal coherence length. 
The author also thanks Roni Harnik, Anson Hook, Shmuel Nussinov, Gilad Perez, and Marianna Safronova for their useful comments. 
The work of AB is supported by NSF grant PHY-2210361 and the
Maryland Center for Fundamental Physics.

	\skipnew{
		\section*{Appendix (To be deleted)}
		For a traditional \Moss set up, a commercially available source of \ce{^{57}Co} with a wavelength of $\mathcal{O}(\AA)$, an activity of $100$ mCi, an initial collimation of $\pm 10^{\circ}$ having a beam width of $w_0=0.1\,{\rm cm}$, results in a photon flux of 
		\bea
		\!\!\!\!\!\!\!\!\!\!\!\!\!\!\!\!\!\!
		N_\gamma \sim 10^4 \left(\frac{10^2\,{\rm }}{L} \frac{w_0}{30\,{\mu\rm m}} \frac{0.03\, {\rm sr}}{\theta_{\rm in}}\right)^2\,\,, 
		\eea
		per month, at the position of the receiver which is $1$ km apart from the source. 
		The Rayleigh range can be calculated to be $\sim 20$ m with the above parameters. 
		To fulfil our requirement of having a \Moss set-up with an almost perfect collimated light beam having tiny to none disperse with distance, we consider existing 
		Existing 
		
		away from the Rayleigh range of the beam with $\theta_{\rm in}$ being the initial beam angular spread~\cite{laser}.

		assumptions of our proposed method is that one can obtain a very high flux of gamma raysphoto
		main experimental challenges of the  stems from obtaining 
		comes from obtaining 
		
		\AB{till here}

		To achieve distances more than micron scale, we need to consider \Moss setup with synchrotron radiation. 
		A synchrotron beam remains coherent over the Rayleigh range of the light~\cite{}, which turns out to be $\mathcal{O}(10^3)\,{\rm km}$ for a beam of width $\mathcal{O}(\rm cm)$ and . 
		
		{\bf Rayleigh Range: } As stated before, in the synchrotron based \Moss spectroscopy, the radiated x-ray after the synchrotron excitation is also coherent. However the Rayleigh range can be calculated as,
		\bea
		z_R = \frac{R_{\rm eff}^2}{\lambda} = \frac{2\pi}{E_\gamma}
		\eea
		
		Index of refraction based mdoel of SMS~\cite{}. 
		
		\subsection{Problems with synchrotron}  
		
		As the synchrotron radiation 
		
		Now we want to estimate the 
		
		Using the SMS, 
		
		Theory of synchrotron~\cite{Hannon1999-ko, Kagan1999-uy}

		The main limitation of the the excitation by synchrotron radiation derives from the achievable statistics.
		This is because of the substantial mismatch between the
		spectral density of the synchrotron radiation and the exceedingly sharp absorption line of the isomeric transitions.
		For a distance, $L$, much smaller than the DM oscillation time, $t_{\rm osc}\sim 1/m_\phi$, the change in the DM field amplitude becomes independent of the   
		
		Note that, if the distance between the Mossbauer emitter and the absorber is $L$ which lights travel in the $\Delta t$ amount of time, then the DM induced change can be written as, 
		\bea
		\Delta \phi \simeq 2 \frac{\sqrt{2 \rho_{\rm DM}}}{m_\phi}\sin\left(\frac{m_\phi\,L}{2}\right) = \sqrt{2 \rho_{\rm DM}}\, L \,\frac{\sin\left(m_\phi\,L/2\right)}{\left(m_\phi\,L/2\right)}\,.
		\label{Eq:del_phi}
		\eea

		are sufficiently far apart, then ultralight DM would induced  
		
		As discussed above, in the \Moss spectroscopy, the small   
		
		and of the nuclear energy levels. 
		
		the nuclear energy 
		
		the nuclear energy levels in the Standard Model (SM), the time oscillation nuclear energy levels would be time-dependent  
		
		f the DM interacts with t 
		
		The idea is to extend the proposal presented in~\cite{Gratta:2020hyp}. 
		The basic idea is that, we have a Mossbauer emitter and a receiver L distance apart. The frequency of the emitter would change in the presence of a time dependent dark matter (DM) background which interacts with the nuclear parameters such as the mass of the nucleon. 
		\\

		\AB{plots of the existing part -- check the number for $\Delta E$ and $N_\gamma$, Study of sychnotron excitation vs traditional mossbauer}

		If we consider the scalar field accounts for the present day dark matter (DM), then the background of $\phi(t,\vec x)$ can be written as,
		\bea
		\phi(t,\vec x)=\frac{\sqrt{2 \rho_{\rm DM}}}{m_\phi}\cos\left[m_\phi (t+\vec\beta\cdot\vec x)\right]\,,
		\eea
		with $\rho_{\rm DM}=0.4\GeV/(\rm cm)^3$ and $|\beta|\sim 10^{-3}$ being the local DM density and the DM virial velocity respectively. 
		The time oscillation of $\phi$ would induce a time-dependence of the nuclear energy levels. 
		For a time interval of $\Delta t$, the time variation of the energy level change in the presence of DM would be proportioal to the change of the DM field amplitude as,
		\bea
		\Delta \phi = \phi(t_i+\Delta t)-\phi(t_i)\simeq 2 \frac{\sqrt{2 \rho_{\rm DM}}}{m_\phi}\sin\left(\frac{m_\phi\,\Delta t}{2}\right)\,,
		\eea
		where, we have omitted a factor of $\sin(m_\phi t_i)\sim \mathcal{O}(1)$. 
		Note that, if the distance between the Mossbauer emitter and the absorber is $L$ which lights travel in the $\Delta t$ amount of time, then the DM induced change can be written as,

		is $L$,between the \Moss emitter and the absorber and its oscillating background, D
		
		M induces changes to the nuclear transition energy at two different times. 
		This would amount to

		due to its ,

		a small net shift in the nuclear transition energy between them. 
		As the

		Thus, even a tiny change in the emitted energy caused by the DM would induce a large deviation of the absorption spectrum.

	}
	
	\newpage
	\bibliography{ref.bib}
\end{document}